\documentclass[conference,compsoc]{IEEEtran}

\ifCLASSOPTIONcompsoc
  \usepackage[nocompress]{cite}
\usepackage{multirow}
\usepackage{wrapfig}
\usepackage{pifont}
\usepackage{graphicx}
\usepackage{amsmath,amsfonts}
\usepackage{algorithm}
\usepackage{algpseudocode}
\usepackage{caption}
\usepackage{graphicx}
\usepackage{multirow}
\usepackage{textcomp}
\usepackage{xcolor}
\usepackage{subfigure}
\usepackage{subcaption}
\usepackage{subfig}

\else
  \usepackage{cite}
\fi

\ifCLASSINFOpdf
\else
\fi

\begin{document}

\title{Smart Grid Security: A Verified Deep Reinforcement Learning Framework to Counter Cyber-Physical Attacks}

\author{
    \IEEEauthorblockN{Suman Maiti, Soumyajit Dey}
    \IEEEauthorblockA{
        Department of Computer Science and Engineering, Indian Institute of Technology, Kharagpur, India
    }
}

\maketitle

\begin{abstract}

The distributed nature of smart grids, combined with sophisticated sensors, control algorithms, and data collection facilities at \emph{Supervisory Control and Data Acquisition} (SCADA) centers, makes them vulnerable to strategically crafted cyber-physical attacks. These malicious attacks can manipulate power demands using high-wattage \emph{Internet of Things} (IoT) botnet devices, such as refrigerators and air conditioners, or introduce false values into transmission line power flow sensor readings. Consequently, grids experience blackouts and high power flow oscillations. Existing grid protection mechanisms, originally designed to tackle natural faults in transmission lines and generator outages, are ineffective against such intelligently crafted attacks. This is because grid operators overlook potential scenarios of cyber-physical attacks during their design phase. 
\par In this work, we propose a safe \emph{Deep Reinforcement Learning} (DRL)-based framework for mitigating attacks on smart grids. The DRL agent effectively neutralizes cyber-physical attacks on grid surfaces by triggering appropriate sequences of existing protection schemes. The safety of the DRL agent is formally verified through a \emph{reachability analysis} method. Additionally, our framework is designed for deployment on CUDA-enabled GPU systems, which enables faster execution of these protection sequences and their real-time validation. Our framework establishes a new set of protection rules for grid models, successfully thwarting existing cyber-physical attacks.

\end{abstract}


%
\IEEEpeerreviewmaketitle

\section{Introduction}

Modern smart grids utilize advanced sensors spread across vast regions and real-time communication methods to distribute electricity efficiently \cite{goel2015security, liu2013framework}. These grids utilize software controls, intelligent sensing, and connectivity to ensure sustainable operations. However, their distributed nature also renders them susceptible to malicious attacks. Attackers can manipulate sensors or remote components in specific regions to distort power flow measurements. Such attacks can lead to unfair power scheduling, creating imbalances in power supply and demand and resulting in blackouts for critical facilities \cite{liang2016review, liang20162015}. Incidents of such attacks have caused significant damage to consumer appliances and infrastructure. For example, in 2015, a coordinated \emph{Load Alteration Attack} (LAA) on Ukrainian \emph{Load Dispatch Centers} (LDCs) led to widespread outages affecting thousands of customers. Additionally, sudden shifts in power demand, caused by the simultaneous on-off sequences of high-wattage IoT devices \cite{soltan2018blackiot, shekari2022madiot}, can disrupt system frequency and lead to cascading failures. Coordinated cyber-physical attacks pose a major threat to modern power grids, targeting sensors, actuators, and communication channels used for control and monitoring. Common types of cyber-physical attacks include \emph{False Data Injection Attacks} (FDIAs) and LAAs, which involve falsifying sensor data and manipulating grid-connected loads. Such attacks can disrupt \emph{Automatic Generation Control} (AGC) units and have been extensively studied for their impact on grid stability in \cite{deka2015one, he2020coordinated, tan2016optimal, anwar2016stealthy, law2014security}.

Existing research on grid security \cite{tan2017modeling,tan2016optimal,tan2013impact} suggests incorporating \emph{Bad Data Detectors} (BDDs) into grid models to detect attack vectors that lead to anomalous grid behavior. However, these BDD units consider only a single state of the grid (operational frequency) and neglect transient dynamics during their design phase, rendering them ineffective against carefully crafted False Data Injection Attacks (FDIAs) \cite{huang2018algorithmic}. Researchers of \cite{huang2019not, ospina2020feasibility} claim that existing grid protection mechanisms can mitigate Load Altering Attacks (LAAs), requiring adversaries to access a vast number of high-wattage IoT botnets to compromise grid behavior. The work in \cite{singer2023shedding} suggests that the grid models used in existing research on grid vulnerability analysis have design flaws due to the absence of crucial control and protection scheme implementations, making the synthesis of attack vectors easier. Contradicting the claims made in \cite{huang2019not, ospina2020feasibility}, and \cite{singer2023shedding}, researchers in \cite{shekari2022madiot} demonstrated that strategically selecting grid buses with the lowest voltage stability index requires significantly fewer botnets to render the grid unsafe, even with protection schemes. In \cite{maiti2023targeted}, a Deep Reinforcement Learning (DRL) agent was utilized to craft FDIA vectors on power flow sensor measurements, successfully compromising standard IEEE bus models despite the presence of AGCs and protection mechanisms. Notably, the attack vectors in this case remained undetected by the BDD units incorporated into the grid model. These findings imply that existing grid protection strategies \cite{huang2019not} are ineffective against state-of-the-art intelligent attack synthesis frameworks that strategically launch attacks on grid surfaces. This underperformance can be attributed to three main reasons. \textit{(i)} Existing protection mechanisms entail significant delays before activation. This delay arises from a hard-bound predefined threshold based on the grid's operating frequency, beyond which protection schemes are triggered with a fixed time delay \cite{huang2019not}. \textit{(ii)} The action space of the protection schemes is limited; they can only shed a preset amount of load or trip all the generators. \textit{(iii)} Grid protection mechanisms are not formulated with sophisticated cyber-physical attacks in mind.

\begin{figure}[!ht]
\includegraphics[width = \columnwidth]{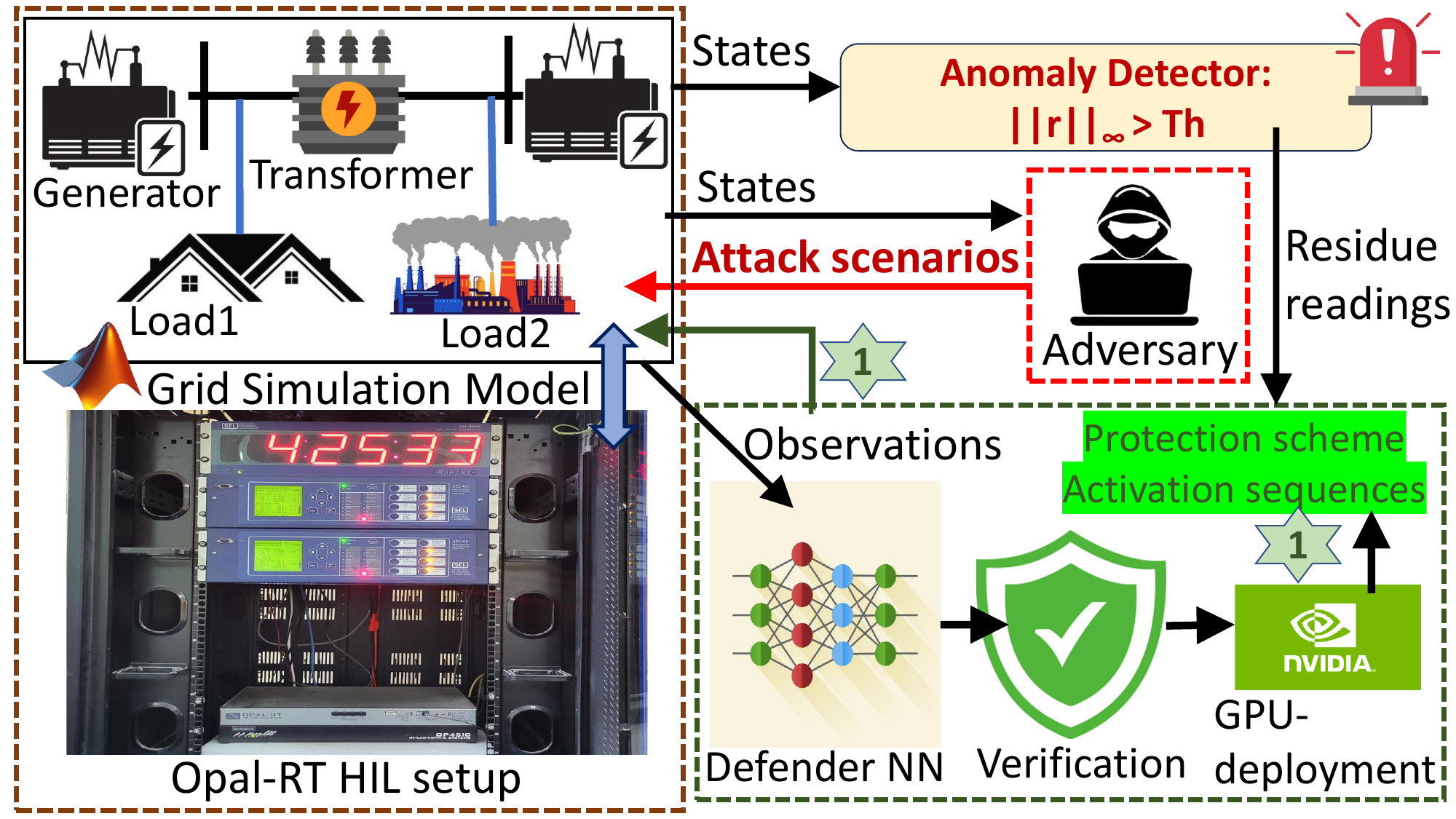}
\caption{Integrated framework to mitigate cyber-physical attacks against smart grids with real-time visualization and evaluation}
\label{Frame}
\end{figure}

Our work addresses the limitations of existing protection mechanisms by adaptively adjusting their activation time and action space according to the attack scenarios. This is accomplished by utilizing a trained DRL \emph{defender} agent which generates effective activation sequences for the existing protection schemes and prevents the grid from blackout and damage to customer equipment \cite{huang2019not}. However, due to the Reinforcement Learning model's black-box nature, the defender agent's actions can be somewhat unpredictable. To address this, it is crucial to formally validate the agent to ensure that it operates safely on deployment in smart grid environments. Therefore, we conduct a reachability analysis on the trained neural network for the agent to check whether there is any overlap with conditions that could lead to unsafe grid operations. Furthermore, our work introduces transient-sensitive anomaly detection units, which are integrated with the generator units within the grid model. These detection units are designed to consider all the grid states. The defender agent also incorporates the responses of these detector units to plan its action sequences, aiming to mitigate the attack's impact quickly. An overview of our grid protection framework is depicted in Fig. \ref{Frame}. 
In summary, our contributions are as follows:

\par \noindent {\bf(1)} We develop novel anomaly detection units for low level monitoring and attack detection in smart grids. Our generalized norm-based detection units take into account all system states and are sensitive to transient dynamics, unlike those in \cite{amini2017hierarchical, tan2016optimal, tan2017modeling,maiti2023targeted}, rendering them more suitable for monitoring grid attacks.

\par \noindent {\bf(2)} We synthesize a DRL based \emph{defender} agent capable of mitigating any cyber-physical attack on smart grids by run time activation of suitable protection scheme action  sequences. The agent is trained using a vast library of attack scenarios across different grid models. Additionally, residue readings from detection units are fed as observations to the defender agent, enhancing its performance in attack scenarios.

\par \noindent {\bf(3)} As smart grids are safety-critical Cyber-Physical Systems (CPS), it is essential to ensure that the DRL induced run time  grid protection  actions are safe. Hence, we formally verify the safety of the defender agent by conducting a suitable formal reachability analysis before deployment. 

\par \noindent {\bf(4)} We demonstrate the practical effectiveness of our methodology through real time hardware-in-the-loop (HIL) emulation of standard IEEE (14, 37, and 39) bus models popularly used in smart grid architectures. The grid models are simulated in-the-loop with our defender agent running on a GPU system for real time validation. 
 The system demonstrates impressive mitigation ability in presence of well known grid attack \cite{shekari2022madiot,maiti2023targeted}.

\textbf{Organization:} The organization of this paper is as follows: Section \ref{Pre} provides essential background information on power grid systems. Section \ref{attack} discusses the state-of-the-art attack models against grid systems. In Section \ref{detector}, we present the mathematical model of the anomaly detection unit. Section \ref{attack mitigation framework} describes our attack mitigation framework. Section \ref{methodology} outlines our methodology, while Section \ref{Experiment} presents the experimental results. The related works are discussed in Section \ref{related}, and the limitations of our framework are explored in Section \ref{limitations}. Finally, Section \ref{future} concludes the paper and discusses future work.

\section{Preliminaries}
\label{Pre}

In this section we provide an overview of the fundamental components of smart grids, focusing on their structure, management, and real-time control mechanisms.

\subsection{Grid Model}
\label{GM}
Smart grids primarily consist of three components: \textit{\textbf{(i) Generation,}} which encompasses generator units converting mechanical power input into electrical power output to meet users' electricity demands; \textit{\textbf{(ii) Transmission,}} where the electrical power produced is transferred to the Load Dispatch Centre (LDC) within each grid region through high-voltage transmission lines; and \textit{\textbf{(iii) Distribution,}} whereby the transmitted power at the LDC is disseminated to industrial and domestic loads via distribution lines. Generation units typically produce power ranging in the hundreds of megawatts. Standard power grid models such as the IEEE 14, 37, and 39 bus configurations feature buses categorized into three types: slack bus, PQ bus, and PV bus. The slack bus establishes an angular reference for all other buses, while the PQ bus determines the voltage magnitude and phase angle of all buses. PV buses are connected to the generators and maintain constant power and voltage generation \cite{kundur2022power}.

\subsection{Grid Operation}
\label{operation}

In smart grids, each generator unit is connected to an Automatic Generation Control (AGC) unit. These AGCs play a vital role in real-time by adjusting the mechanical power set-points of the generators according to the power demand in the grid \cite{kundur2022power}. This adjustment is crucial for maintaining the grid frequency at its designated value within acceptable boundaries. Furthermore, AGCs are responsible for managing the power exchange between grid areas, ensuring that the total power transmitted over tie lines stays at a predetermined level with the help of a metric named \emph{Area Control Error} (ACE).

\begin{figure}[!ht]
\includegraphics[width = \columnwidth]{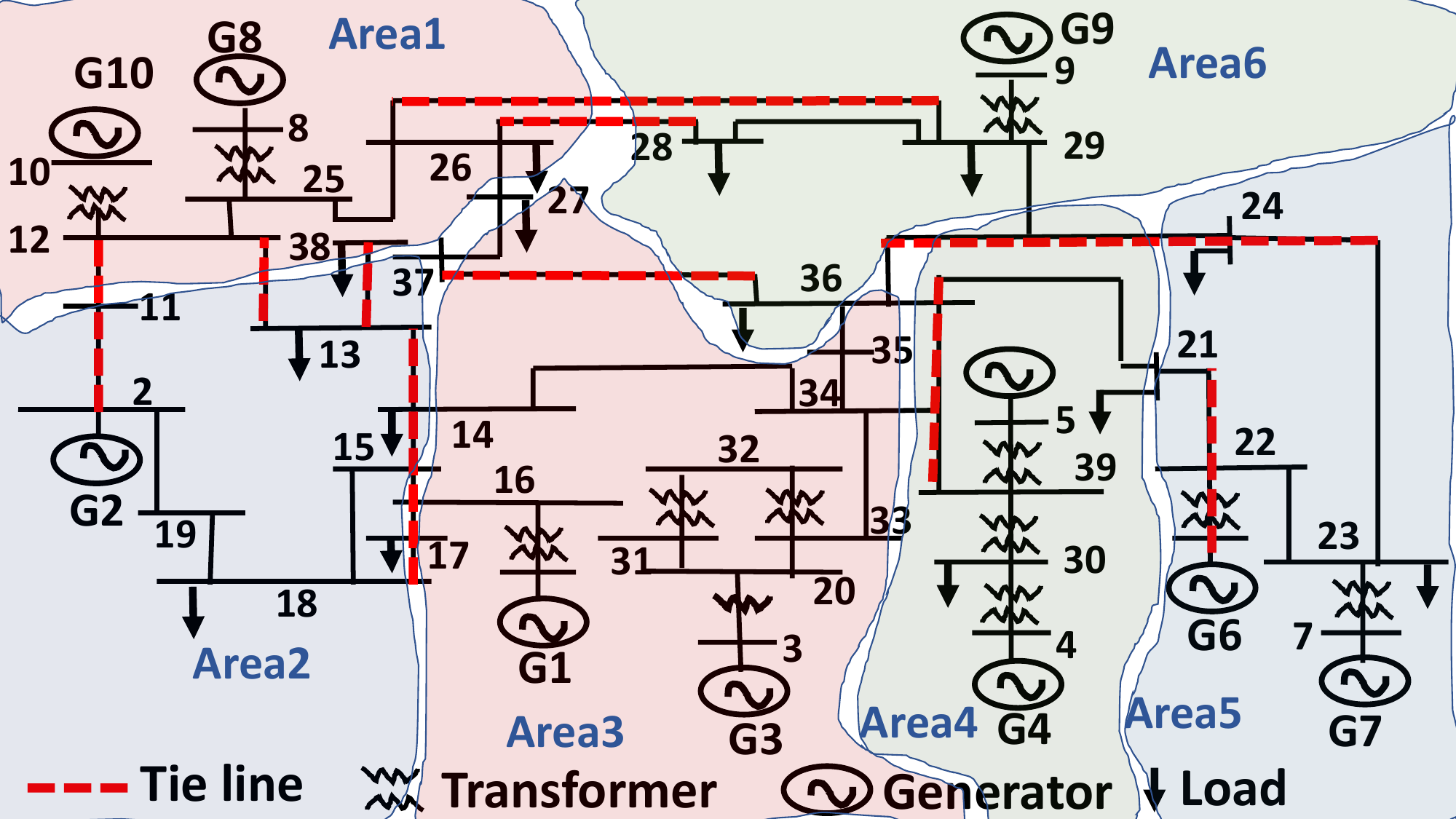}
\caption{IEEE 39 bus power grid with generators, transmission and distribution lines \cite{maiti2023targeted}.}
\label{fig:AGCarea}
\end{figure}

A tie-line connects two buses in separate areas, as illustrated in Fig.\ref{fig:AGCarea}, which shows a grid with six areas and 39 buses, with tie-lines indicated by red dotted lines. To calculate the ACE for the $n^{th}$ tie-line at time $t$, the AGC measures deviations of the grid frequency and power export from their nominal values as ${\Delta f}_{n}(t)$ and $\Delta{p_e}_{n}(t)$, respectively. Using these measurements, AGC calculates ACE as $ACE_{n}(t) =\alpha_{n}(t)\Delta{p_e}_{n}(t)+\beta_{n}(t)\Delta f_{n}(t)$, where $\alpha_{n}(t)$ and $\beta_{n}(t)$ are time-varying constants. The ACE value is updated once every AGC cycle (generally of length six to ten seconds)~\cite{mohanty2016design}, and accordingly, the mechanical input power set points of the generators are updated to keep the ACE value near zero \cite{obaid2019frequency}. To ensure safe grid operation, the grid frequency deviation ${\Delta f}$ from its nominal value should always remain within specific safety limit $S=[-0.5,0.5]$Hz (\cite{obaid2019frequency}).

\subsection{Protection Schemes}
\label{protection}

During natural events such as short circuits in transformers or sudden transmission line outages due to storms and rain, grids rely on power system protection schemes to maintain safe operation. These schemes aim to swiftly isolate the affected area and minimize damage to grid components. By implementing localized outages, these protection schemes facilitate repairs while ensuring the majority of the transmission network remains functional. However, these existing strategies are triggered by some time delays by the operators at the LDCs. For clarity, Table \ref{protection table} depicts the existing protection schemes considered in our grid models, which are also widely utilized in bulk power systems \cite{huang2019not, shekari2022madiot, soltan2018blackiot, maiti2023targeted}. The unit specified in the table denotes the per-unit value of the grid quantity, where a value of 1 \emph{pu} indicates that the quantity is at its nominal value \cite{kundur2022power}.

\begin{table}[]
\resizebox{\columnwidth}{!}{%
\begin{tabular}{|l|lll|}
\hline
\textbf{\begin{tabular}[c]{@{}l@{}}Name of \\ protection scheme\end{tabular}} & \multicolumn{3}{c|}{\textbf{Activation criteria}}                               \\ \hline
Distance                                                                & \multicolumn{3}{l|}{Short circuit fault}                                        \\ \hline
Overcurrent                                                             & \multicolumn{3}{l|}{Current = 2 {[}pu{]}}                                       \\ \hline
Under Voltage (UV)                                                      & \multicolumn{3}{l|}{Voltage = 0.8 {[}pu{]}}                                     \\ \hline
\multirow{4}{*}{\begin{tabular}[c]{@{}l@{}}Under Frequency \\ Load Shedding \\ (UFLS) \end{tabular}} &
  \multicolumn{1}{l|}{\textbf{\begin{tabular}[c]{@{}l@{}}Frequency \\ threshold\end{tabular}}} &
  \multicolumn{1}{l|}{\textbf{\begin{tabular}[c]{@{}l@{}}System load\\  relief\end{tabular}}} &
  \textbf{\begin{tabular}[c]{@{}l@{}}Time \\ delay\end{tabular}} \\ \cline{2-4} 
                                                                              & \multicolumn{1}{l|}{59.3 Hz}                & \multicolumn{1}{l|}{5\%}   & 0 Sec      \\ \cline{2-4} 
                                                                              & \multicolumn{1}{l|}{58.9 Hz}                & \multicolumn{1}{l|}{15\%}  & 0 Sec      \\ \cline{2-4} 
                                                                              & \multicolumn{1}{l|}{58.3 Hz}                & \multicolumn{1}{l|}{25\%}  & 0 Sec      \\ \hline
\multirow{4}{*}{\begin{tabular}[c]{@{}l@{}}Over Frequency   \\ Generator Tripping \\ (OFGT) \end{tabular}} &
  \multicolumn{1}{l|}{\textbf{\begin{tabular}[c]{@{}l@{}}Frequency \\ threshold\end{tabular}}} &
  \multicolumn{1}{l|}{\textbf{\begin{tabular}[c]{@{}l@{}}Generation \\ trip\end{tabular}}} &
  \textbf{\begin{tabular}[c]{@{}l@{}}Time \\ delay\end{tabular}} \\ \cline{2-4} 
                                                                              & \multicolumn{1}{l|}{60.6 Hz}                & \multicolumn{1}{l|}{100\%} & 9 min  \\ \cline{2-4} 
                                                                              & \multicolumn{1}{l|}{61.6 Hz}                & \multicolumn{1}{l|}{100\%} & 30 Sec \\ \cline{2-4} 
                                                                              & \multicolumn{1}{l|}{\textgreater = 61.8 Hz} & \multicolumn{1}{l|}{100\%} & 0 Sec  \\ \hline
\multirow{5}{*}{\begin{tabular}[c]{@{}l@{}}Under Frequency   \\ Generator Tripping\\  (UFGT) \end{tabular}} &
  \multicolumn{1}{l|}{\textbf{\begin{tabular}[c]{@{}l@{}}Frequency \\ threshold\end{tabular}}} &
  \multicolumn{1}{l|}{\textbf{\begin{tabular}[c]{@{}l@{}}Generation \\ trip\end{tabular}}} &
  \textbf{\begin{tabular}[c]{@{}l@{}}Time \\ delay\end{tabular}} \\ \cline{2-4} 
                                                                              & \multicolumn{1}{l|}{59.4 Hz}                & \multicolumn{1}{l|}{100\%} & 9 min  \\ \cline{2-4} 
                                                                              & \multicolumn{1}{l|}{58.4 Hz}                & \multicolumn{1}{l|}{100\%} & 30 Sec \\ \cline{2-4} 
                                                                              & \multicolumn{1}{l|}{58.0 Hz}                & \multicolumn{1}{l|}{100\%} & 2 Sec  \\ \cline{2-4} 
                                                                              & \multicolumn{1}{l|}{57.5 Hz}                & \multicolumn{1}{l|}{100\%} & 0 Sec  \\ \hline
Differential                                                            & \multicolumn{3}{l|}{Symmetrical/asymmetrical fault in transmission line}        \\ \hline
\end{tabular}%
}
\caption{Traditional protection schemes used in power grids \cite{maiti2023targeted, huang2019not}.}
\label{protection table}
\end{table}

\section{Attack models}
\label{attack}

\begin{table*}[]
\begin{tabular}{|c|cc|cc|c|c|}
\hline
\multirow{2}{*}{\textbf{\begin{tabular}[c]{@{}c@{}}Attack \\   Model\end{tabular}}} &
  \multicolumn{2}{c|}{\textbf{Attacker   Objective}} &
  \multicolumn{2}{c|}{\textbf{Attacker Capabilities}} &
  \multirow{2}{*}{\textbf{\begin{tabular}[c]{@{}c@{}}Presence of \\ Protection\\ Scheme\end{tabular}}} &
  \multirow{2}{*}{\textbf{\begin{tabular}[c]{@{}c@{}}\# States \\    in \\ Detectors\end{tabular}}} \\ \cline{2-5}
 &
  \multicolumn{1}{c|}{\textbf{Target Components}} &
  \textbf{\begin{tabular}[c]{@{}c@{}}Grid quantity \\ manipulated\end{tabular}} &
  \multicolumn{1}{c|}{\textbf{Attack Resources}} &
  \textbf{Action Space} &
   &
   \\ \hline
Black IoT {\cite{soltan2018blackiot}} &
  \multicolumn{1}{c|}{Consumer Loads} &
  Power Flow &
  \multicolumn{1}{c|}{\begin{tabular}[c]{@{}c@{}}High-wattage \\ IoT devices\end{tabular}} &
  \begin{tabular}[c]{@{}c@{}}Switching\\  on-off \\ devices\\  synchronously\end{tabular} &
  $\checkmark$ &
  NA \\ \hline
MaD IoT {\cite{shekari2022madiot}} &
  \multicolumn{1}{c|}{Consumer Loads} &
  Power Flow &
  \multicolumn{1}{c|}{\begin{tabular}[c]{@{}c@{}}High-wattage\\  IoT devices\end{tabular}} &
  \begin{tabular}[c]{@{}c@{}}Switching \\ on-off \\ devices \\ synchronously\end{tabular} &
  $\checkmark$ &
  NA \\ \hline
FDIA {\cite{tan2016optimal, tan2017modeling}} &
  \multicolumn{1}{c|}{AGC control signal} &
  \begin{tabular}[c]{@{}c@{}}Generator output \\ electrical power\end{tabular} &
  \multicolumn{1}{c|}{\begin{tabular}[c]{@{}c@{}}Grid topology or \\ state information\end{tabular}} &
  Sensors &
 \ding{55}&
  1 \\ \hline
Scaling Attack {\cite{maiti2023targeted}} &
  \multicolumn{1}{c|}{Tie-line power flow measurements} &
  Power Flow &
  \multicolumn{1}{c|}{\begin{tabular}[c]{@{}c@{}}Grid topology or \\ state information\end{tabular}} &
  Sensors &
  $\checkmark$ &
  5 \\ \hline
CBT {\cite{case2016analysis}} &
  \multicolumn{1}{c|}{Transmission/Distribution ines} &
  Power Flow &
  \multicolumn{1}{c|}{LDC circuit breakers} &
  \begin{tabular}[c]{@{}c@{}}Open/close \\ circuit \\ breakers \\ synchronously\end{tabular} &
  $\checkmark$ &
  NA \\ \hline
\end{tabular}
\caption{Cyber-physical attacks on smart grids}
\label{CPS attacks}
\end{table*}

\begin{table}[]
\begin{tabular}{|c|c|c|}
\hline
\textbf{Sl.No} & \textbf{State Name}                                                                            & \textbf{State Symbol} \\ \hline
1  & Rotor angle                    & $ \theta $          \\ \hline
2  & Flux at phase a                & $ \phi_{a} $        \\ \hline
3  & Flux at phase b                & $ \phi_{b} $        \\ \hline
4  & Flux at phase c                & $ \phi_{c} $        \\ \hline
5  & Direct - axis flux             & $ \phi_{d} $        \\ \hline
6  & Quadrature - axis flux         & $ \phi_{q} $        \\ \hline
7  & Current at phase a             & $ I_{a} $           \\ \hline
8  & Current at phase b             & $ I_{b} $           \\ \hline
9  & Current at phase c             & $ I_{c} $           \\ \hline
10 & Direct - axis current          & $ I_{d} $           \\ \hline
11 & Quadrature - axis current      & $ I_{q} $           \\ \hline
12 & Field Current                  & $ I_{fd} $          \\ \hline
13             & \begin{tabular}[c]{@{}l@{}}Time delay for rotor to respond \\ to mechanical power\end{tabular} & $d\omega_{delay}$     \\ \hline
14 & Predicted rotor speed delay    & $d\omega_{predict}$ \\ \hline
15 & Voltage at phase a             & $V_{a}$             \\ \hline
16 & Voltage at phase b             & $V_{b}$             \\ \hline
17 & Voltage at phase c             & $V_{c}$             \\ \hline
18 & Generator operation frequency                    & $f$            \\ \hline
19 & Saturated inductance at d-axis & $L_{md sat}$ \\ \hline
\end{tabular}
\caption{Grid states}
\label{States considered}
\end{table}

In this section, we discuss various types of cyber-physical attack models designed in existing literature to compromise the safety of smart grid operations. We categorize these models according to their target components, the grid quantities they manipulate, their required resources, and their action space. We detail the features of these attack models in Table \ref{CPS attacks}, along with the indication of whether protection mechanisms and detector units were incorporated into the grid models in these works. The term NA indicates the absence of a detector in the grid model. In this work, we focus on attack models that directly affect power flow distribution within the grids. Other threat models, such as market manipulation \cite{shekari2021mamiot}, are beyond the scope of this paper. Below are the attack models considered in this work.

\noindent \textbf{Black IoT \cite{soltan2018blackiot}:} In this attack framework an adversary first gains control of high-wattage IoT-based devices such as heaters, air conditioners, and refrigerators. It then synchronously switches these botnet devices on and off until the grid frequency deviates greatly from its safe range ($S$) of [59.5 60.5] Hz. It was claimed in \cite{soltan2018blackiot} that only a 1\% increase in load demand with the help of botnets can cause a blackout in the Polish grid. These attacks are also known as \emph{Manipulation of Demand via IoT} (MaD IoT 1) attacks.

\noindent \textbf{MaD IoT 2.0\cite{shekari2022madiot}:} This is a modified version of the MaD IoT 1 attack where the attacker first computes the voltage stability index of each bus node in a grid model by performing a modal analysis \cite{kundur2022power} of the grid. The attacker subsequently alters the loads connected to the buses with the least voltage stability index using high-wattage IoT botnet devices to make grid operation unsafe. For clarity purposes, we will refer to the MaD IoT 1 attack as Black IoT and MaD IoT 2.0 as MaD IoT for the rest of the paper.  

\noindent \textbf{False Data Injection Attack (FDIA) \cite{tan2016optimal, tan2017modeling}:} In these types of attacks, the adversary introduces false readings to the generators' reference power measurements without triggering the BDD units. The aim of the attacker in these works is to create a mismatch between power supply and demand.

\noindent \textbf{Scaling Attacks  \cite{maiti2023targeted}:} In this attack model, the adversary first identifies the most vulnerable operating instances of a grid with the help of formal analysis \cite{abbas2013probabilistic}. It then subsequently scales the power flow sensor measurements of the tie lines during the identified intervals, which leads to increased ACE values and makes grid operation unsafe. Such a type of attack also manages to remain undetected by sophisticated transient-sensitive anomaly detectors.

\noindent \textbf{Circuit Breaker Takeover (CBT) \cite{case2016analysis}:} An attacker compromises the LDCs by social engineering \cite{tan2016optimal} and disconnects loads, generators, and transmission lines from the grid using relay/circuit break-on-off sequences. This action leads to power flow oscillations within the grid.

We implemented the discussed attack scenarios in our grid models to effectively train the DRL defender agent. We assume that these adversaries have read/write access to three types of resources: $(i)$ power flow sensor measurements of the tie lines ($T_{PF}$) \cite{maiti2023targeted}, $(ii)$ circuit breaker control signals ($CB$) \cite{case2016analysis}, and $(iii)$ generator reference power $P_{ref}$ measurements. We consider that the ultimate objective of the attack models is to inflict maximum damage on consumer equipment by pushing the grid frequency ($f$) beyond the safe equipment operation zone, defined here as $D \in [58, 62]$ Hz \cite{soltan2018blackiot} across various grid topologies.

\section{Anomaly Detection in Smart Grid}
\label{detector} 
\begin{wrapfigure}{r}{0.3\textwidth}
  \centering
  \includegraphics[width=0.3\textwidth]{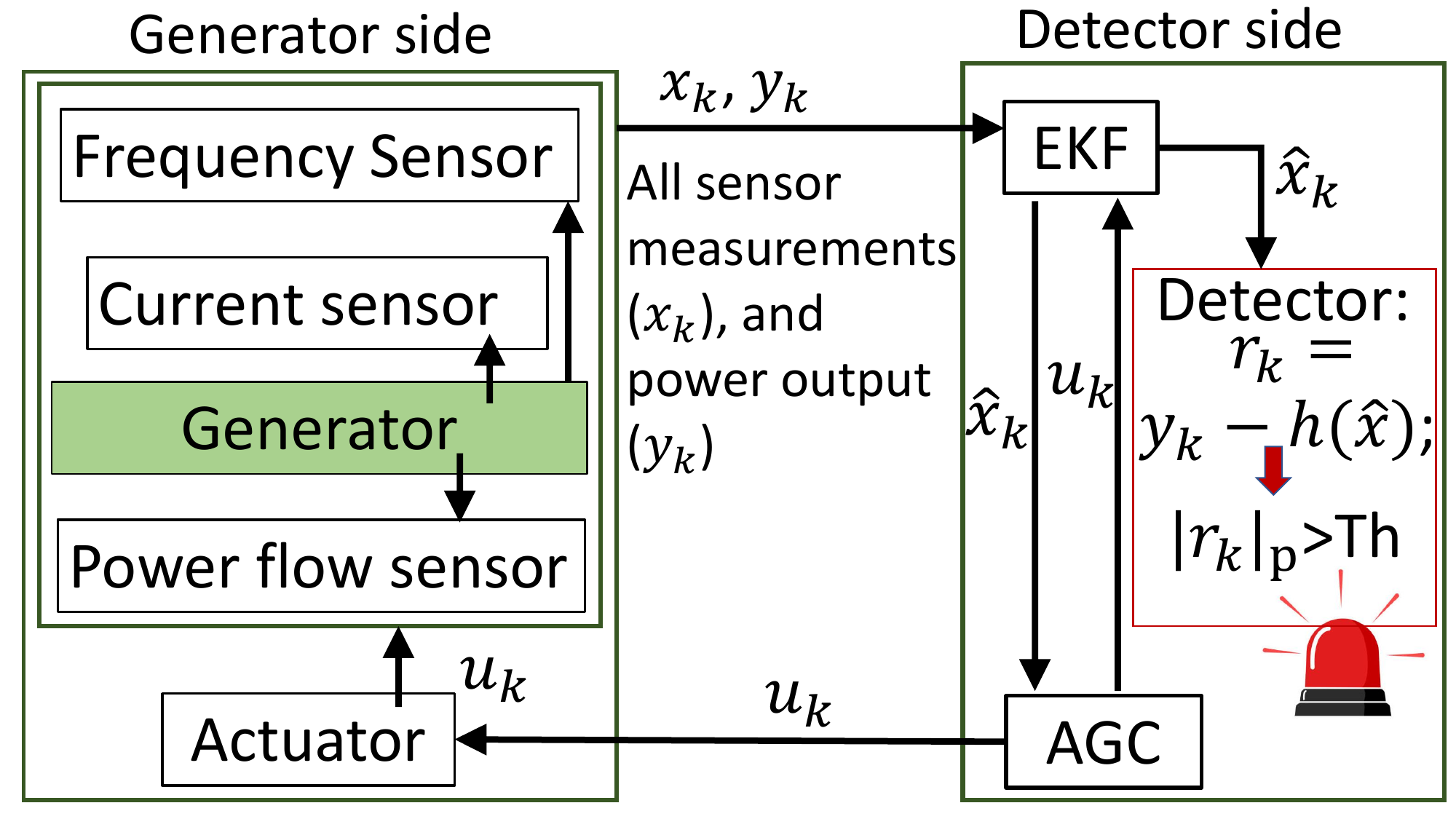}
  \caption{Anomaly detection framework}
  \label{control framework}
\end{wrapfigure}

In this work, we develop novel anomaly detection units that are attached to each generator unit in the grid for discovering deviations in their behavior from nominal operations. The operational framework of the anomaly detector is demonstrated in Fig. \ref{control framework}. These detectors utilize an \textit{Extended Kalman Filter} (EKF) \cite{ribeiro2004kalman} based state estimator to estimate generator system state $\hat{x}$ as given by, $\hat{x}_{k+1} = g(\hat{x}_k, u_k) + L(y_k - h(\hat{x}_k))$ where \(g\) represents the state transition function \cite{fujii2013extended}, \(h\) is the measurement function \cite{fujii2013extended}, \(L\) is the Kalman gain, \(u_k\) is the control input (mechanical input power set point of generator), and \(y_k\) is the output power measurement of the generator at the \(k\)-th time step.

Thereafter, the residue $(r_k)$, defined by, $r_k = y_k - h(\hat{x}_k)$ is utilized to detect deviations of the generator system states from their nominal behavior. As an anomaly condition, we check if the norm of the system residue exceeds a predetermined threshold $(Th)$. Let $|r_{k}|_p$ denote the $p$-th norm of the residue at the $k$-th time stamp. Considering that residue $r_k$ has dimension $n_r$, let $r_k=[ {r^1_k}^T,\cdots , {r^{n_r}_k}^T]^T$. The residue is calculated over a $w$ length window as follows. $|r_k|_p \overset{\Delta}{=} \left( \sum_{i=1}^w \sum_{j=1}^{n_r} |r_{i}^j|^p \right)^{1/p} > Th$. Previous research works \cite{tan2013impact, tan2016optimal, tan2017modeling} have focused primarily on using grid frequency ($f$) as the sole state parameter for designing anomaly detectors. This approach often renders the detectors insensitive to the transient dynamics of the grid. The work in, \cite{maiti2023targeted} extends the state estimation by considering five state variables, employing the Kalman Filter (KF) as the estimator. However, the use of KF often acts as a bottleneck when estimating the nonlinear dynamics of the grid, thus failing to effectively detect anomalous grid behavior. In contrast, our anomaly detector incorporates a broader spectrum of grid states, specifically nineteen states as detailed in Table \ref{States considered}. By utilizing the Extended Kalman Filter (EKF) for state estimation, our system is finely attuned to the transient dynamics of the grid, significantly enhancing its efficacy in detecting grid anomalies.

\section{Attack Mitigation Framework Overview}
\label{attack mitigation framework}

\begin{figure}[!ht]
\includegraphics[width = \columnwidth]{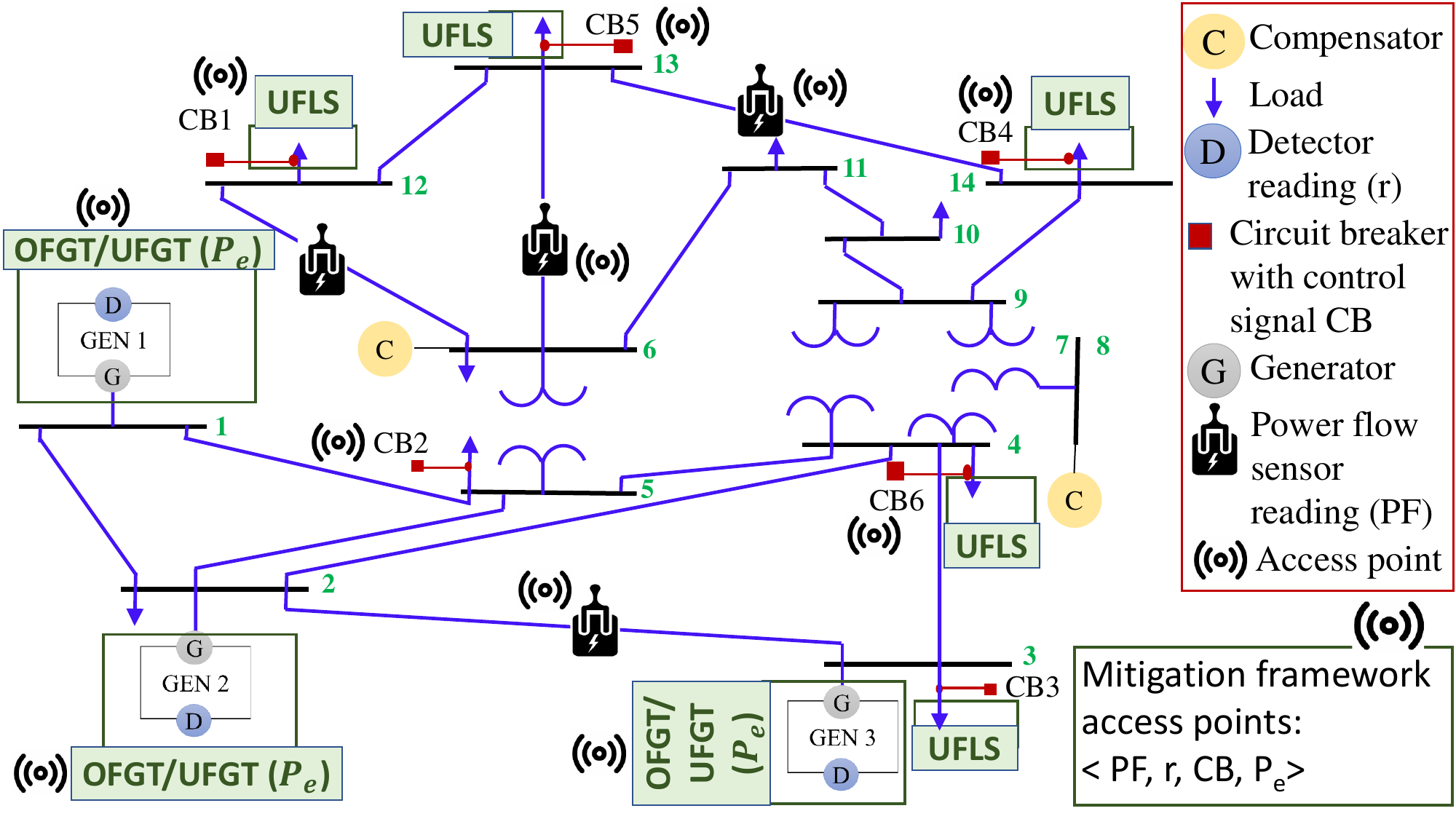}
\caption{IEEE 14-bus power grid with all existing protection schemes and the access points of the mitigation framework incorporated.}
\label{mitigation framework}
\end{figure}

We incorporate all existing protection mechanisms listed in Table \ref{protection table} within our grid models. Fig.\ref{mitigation framework} illustrates the locations of different protection schemes within an IEEE 14-bus grid model utilized in our work. Our attack-mitigation framework is capable of sequentially triggering these protection schemes based on various attack scenarios, as it has read/write access to the following: $(i)$ all the transmission line power flow sensor measurements ($PF$), $(ii)$ the residue readings from the anomaly detection units ($r$), $(iii)$ relay and circuit breaker control signals ($CB$), and $(iv)$ the ability to control the power output of each generator unit ($P_e$). We represent these access points in the form of a tuple $\langle PF, r_{ad}, CB, P_e \rangle$, as depicted in Fig.\ref{mitigation framework}. Utilizing these access points, our mitigation framework can adaptively adjust the activation of the existing protection schemes instantly according to the specific attack scenarios. For example, using $CB$ signals, our framework can trigger Under Frequency Load Shedding (UFLS) by disconnecting loads from the grid, and with $P_e$, it activates Over Frequency Generator Tripping (OFGT) and Under Frequency Generator Tripping (UFGT) by reducing the power production of generators. The combined use of $PF$, $r$, and $CB$ enables the framework to activate overcurrent, under-voltage, and differential protection schemes by disconnecting certain transmission and distribution lines as needed. The framework's capabilities are structured to ensure immediate response to various attack scenarios, making it suitable for use in Load Dispatch Centers (LDCs) where grid operators have similar access controls.

The attack models discussed in section~\ref{attack} aim to disrupt the grid's power flow, leading to significant frequency deviations ($f$) and potential damage to customer equipment. To stabilize uncontrolled power flow oscillations, our mitigation framework seeks to minimize the difference between the power flow during attack conditions and the power flow during normal grid operations across grid buses. Additionally, it aims to maintain the grid frequency ($f$) within the safe operational range of $S \in [59.5, 60.5]$ Hz, close to the nominal frequency of 60 Hz, by initiating appropriate protection strategy sequences. Let $P_{norm}$ denote the power flow between buses $i$ and $j$ during normal grid operation, and let $P_{atk}$ denote the power flow between the same buses under attack conditions. These power flow values are mathematically described by the following equations.

\begin{equation}
\label{mit1}
P_{norm} = \frac{(V_{S_j}^2 - V_{S_i}^2)}{X_{L_{ji}}} \sin(\delta_j - \delta_i)
\end{equation}

\begin{equation}
\label{mit2}
P_{atk} = \frac{({V_{S_j}^a}^2 - {V_{S_i}^a}^2)}{X_{L_{ji}}} \sin({\delta_j}^a - {\delta_i}^a)
\end{equation}

Here, the magnitudes of the voltages at buses $i$ and $j$ during normal grid operation are denoted by $V_{S_i}$ and $V_{S_j}$, respectively. The voltage angles at these buses are represented by $\delta_i$ and $\delta_j$, respectively, and $X_{L_{ji}}$ represents the line impedance between the two buses. During an attack scenario, the bus voltages are denoted by $V_{S_i}^a$ and $V_{S_j}^a$, and their respective voltage angles are represented by $\delta_i^a$ and $\delta_j^a$. The objectives of our mitigation framework are mathematically described by the following equation.

\begin{equation}
\label{mit3}
    \min_{\langle CB, P_e, PF \rangle} \max_{\theta \in \Theta} \left[ C(P_{norm}, P_{atk_{\theta}}, f) \right] \approx 0
\end{equation}

Here, $C(P_{\text{norm}}, P_{\text{atk}_{\theta}}, f) = w_1 |P_{\text{norm}} - P_{\text{atk}}| + w_2( -1+|\frac{|f|-60}{0.5}|)$, is the cost function that quantifies the deviation in grid behavior under attack conditions and $w_{1},w_{2}$ are the weight values used to prioritize the components of the cost function. $\Theta$ is the set of all potential scenarios, representing different attack vectors, and $\theta$ is a specific scenario from the set $\Theta$. Our framework minimizes the deviation in grid behavior from its normal operation against the worst-case attack scenarios by performing the minimization operation depicted in Equation \ref{mit3} for every possible pair of buses in a grid model. It determines the optimal protection scheme activation sequence $(CB, P_e, PF)$ for every attack scenario. Additionally, we assume that our mitigation framework cannot control certain grid equipment, such as circuit breakers, which have been compromised by the adversary to induce attacks in the grid model.

\subsection{Motivating Example}
\label{example}

\begin{wrapfigure}{r}{0.6\columnwidth}
  \centering
  \includegraphics[width=0.6\columnwidth, clip]{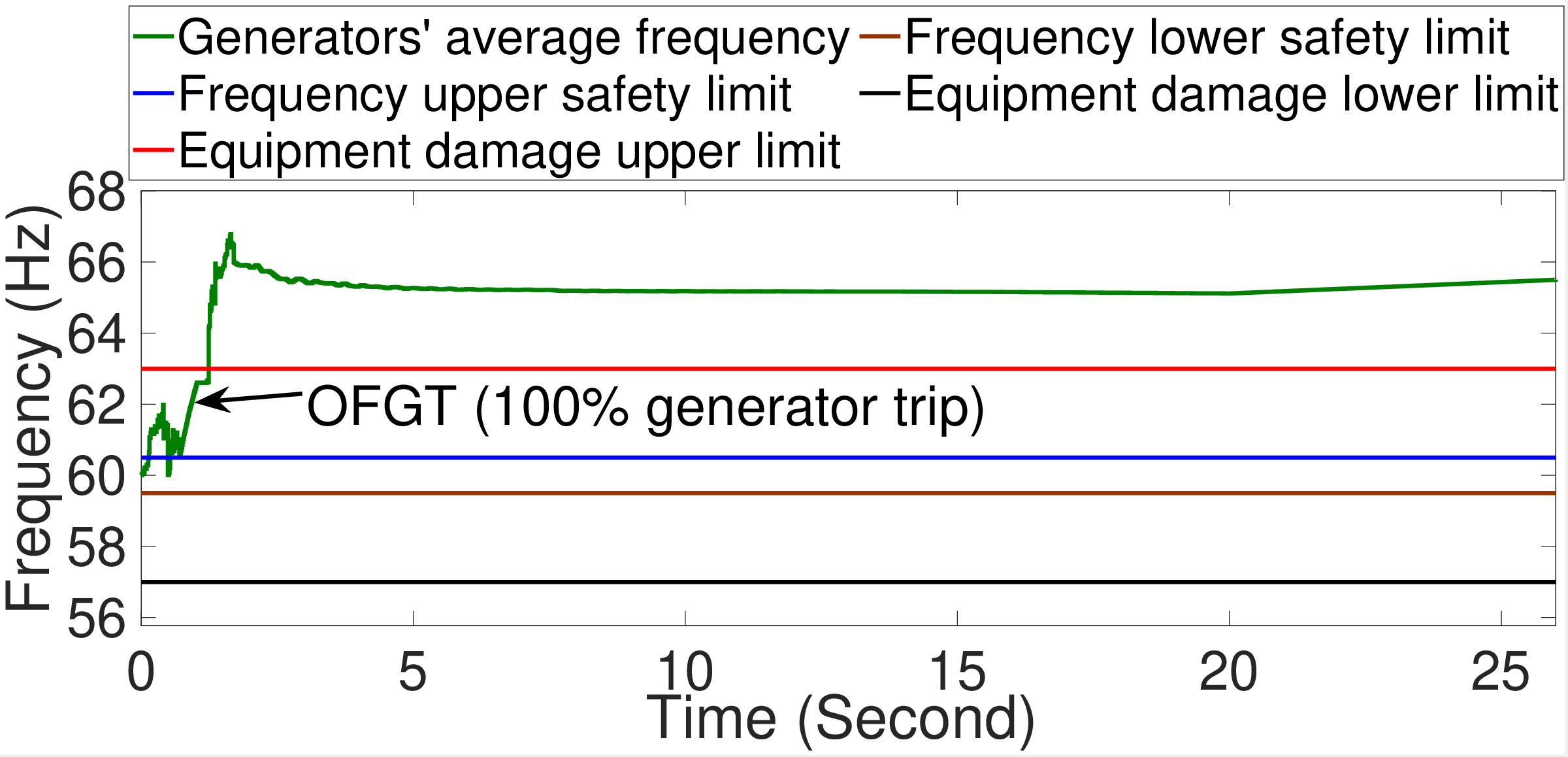}
  \caption{The operating frequency of an IEEE 39-bus grid, assessed under the influence of attack vectors proposed in \cite{maiti2023targeted}, with traditional protection systems engaged.}
  \label{mot1}

  \vspace{6pt} 
 \includegraphics[width=0.6\columnwidth, clip]{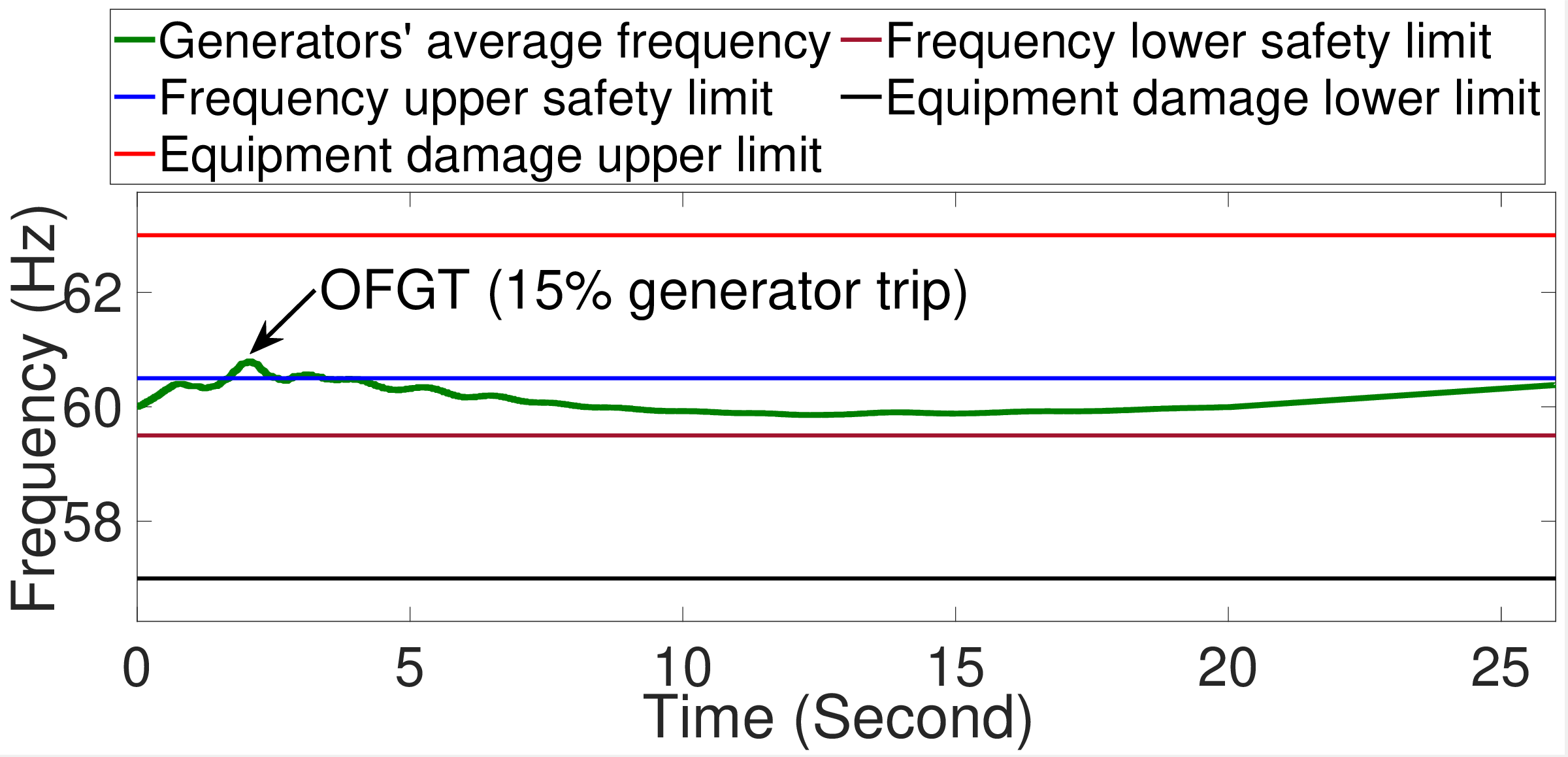}
  \caption{The operating frequency of an IEEE 39-bus grid, assessed under the influence of attack vectors proposed in \cite{maiti2023targeted}, with our mitigation framework in action.}
  \label{mot2}
\end{wrapfigure}

In this section, we demonstrate how our attack mitigation framework successfully prevents grid failure against state-of-the-art attack models. We present an example of how the frequency trajectory of the New England (IEEE 39 bus) grid changes when scaling attacks \cite{maiti2023targeted} on tie-line power flow measurements are implemented. The green plot in Fig.\ref{mot1} illustrates the average operating frequency of all generators in the IEEE 39 bus grid model under the influence of scaling attacks. 

The blue, brown, red, and black plots represent the upper and lower safety limits for operating frequency, as well as the upper and lower safety limits for equipment damage, respectively. The attack pushes the grid operating frequency beyond the safe limit within 1 second. The frequency trajectory continues to deviate and crosses the upper limit for equipment damage $D$ (\cite{soltan2018blackiot}) at 3 seconds. The scaling attack successfully causes damage to customer equipment and jeopardizes grid operation safety. Traditional protection mechanism, triggers the \emph{Over Frequency Generator Triping} (OFGT) method, with 100\% generation trip at 2.5 seconds instead of activating it immediately when the frequency exceeds the safe operation limit ($S$). This delay in the operation of the protection scheme leads to equipment damage for consumers.  We now demonstrate how our attack mitigation framework successfully ensures grid operation safety when subjected to scaling attacks. The green plot in Fig.\ref{mot2} depicts the average operating frequency of generators in an IEEE 39 bus grid model under scaling attacks. Similarly, the blue, brown, red, and black plots denote the upper and lower safety limits for operating frequency, as well as the upper and lower safety limits for equipment damage \cite{soltan2018blackiot}, respectively. As observed in Fig.\ref{mot2}, the grid frequency surpasses the upper safety limit at 1 second. Subsequently, our mitigation framework activates the OFGT scheme with only a 15\% generation trip at 1.1 seconds, thereby constraining the frequency trajectory to return within the safe zone ($S$). This strategy effectively prevents damage to customer equipment and ensures grid operation safety. Immediate activation of an appropriate protection strategy when the operating frequency becomes unsafe prevents customer equipment damage and ensures grid operation safety, thereby avoiding total generation shutdown. This motivates us to develop an adaptive attack mitigation framework capable of strategically activating the protection schemes at appropriate time intervals to ensure grid operation safety and prevent damage to customer equipment against different classes of cyber-physical attacks.

\section{Methodology for Attack Mitigation}
\label{methodology}

We consider as input $(i)$ a smart grid model in Simulink, comprising various loads, generators, transmission, and distribution lines, and $(ii)$ a library of different attack scenarios, including FDIA (\cite{tan2016optimal, tan2017modeling}), scaling (\cite{maiti2023targeted}), CBT (\cite{case2016analysis}) and MaD IoT attacks (\cite{soltan2018blackiot, case2016analysis, shekari2022madiot}). Our methodology for the attack mitigation framework consists of two components. $(i)$ First, we train a Deep Reinforcement Learning (DRL) based defender agent using various attack scenarios on different standard IEEE grid architectures. The RL defender agent learns to stochastically neutralize the effects of attacks on grid behavior by discovering appropriate activation sequences for the protection mechanisms (Table \ref{protection table}). $(ii)$ After completing the training process, we perform a set-based verification of the agent's neural networks (NNs). This verification computes the reachable set of states \cite{bak2017hylaa} of the agent's NNs and checks their intersection with the unsafe grid states. Our proposed methodology outputs formally verified optimal protection scheme activation strategies against state-of-the-art attack models, which are deployable in safety-critical smart grid environments.

\subsection{DRL Defender agent for Grid Protection}
\label{defender}

Following the discussion in section \ref{attack mitigation framework} the defender agent's objective is to make the grid operation safe by optimally triggering a sequence of protection schemes at appropriate time instants such that the cost function in equation \ref{mit3} gets minimized. The defender takes action whenever the input attack vectors make grid operation unsafe and cause equipment damage or when the attack vectors get discovered by the anomaly detection units.

Our defender DRL agent's action space includes:
$(i)$ triggering UFLS, UV, and differential protection by adjusting the circuit breaker control signals ($CB$).
$(ii)$ Activating OFGT and UFGT by changing the generator output power production ($P_e$). For synthesizing the best grid protection activation strategy to make grid operation safe, the defender agent observes the following data from the power grid:
$(i)$ The operating frequency ($f$) of each generator unit in the grid,
$(ii)$ the transmission line power flow sensor measurements ($PF$),
$(iii)$ the residue values of each generator unit ($r_{k}$),
$(iv)$ the threshold values of anomaly detection units ($Th$),
$(v)$ the circuit breaker control signals ($CB$), and
$(vi)$ the output power produced by the generators ($P_e$). Equation \ref{rwddef} as given below formulates the reward function where $w_1'$, $w_2'$, and $w_3'$ are weight parameters.

{\scriptsize
\begin{equation}
\begin{aligned}
R_{d} = & \, -\underbrace{w'_1\sum_{k=0}^{d-1} \left( \sum_{i=0}^{n} -1+\left|\frac{|f_{i}|-60}{0.5}\right|\right)}_{\text{minimize frequency deviation}} \\
& - \underbrace{w'_2\sum_{m=1}^{b}\sum_{q=1}^{b} |P_{\text{norm}_{mq}} - P_{\text{atk}_{mq}}|}_{\text{minimize power flow deviation}} \\
& - \underbrace{w'_3\sum_{k=0}^{d-1}\sum_{i=1}^{n} \left( \frac{Th_{i}-2 \max(0, |r_{ik}|_{p} - Th_i)}{Th_i}\right)}_{\text{minimize system residue}}
\end{aligned}
\label{rwddef}
\end{equation}
}

There are three components of the reward function designed to ensure grid safety and prevent damage to customer equipment. The first component is maximized when the frequency deviation of all $n$ generators in the grid over $d$ sampling windows remains within the safe operating range ($S \in$ [59.5, 60.5] Hz). The second component of the reward function is maximized when the power flow deviations between two buses $m$ and $q$ in a $b$-bus grid are minimized. The third component is maximized when, over a $d$-length time horizon, the $p$-th norm of the system residue remains below the threshold ($Th$). A larger residue value ($|r_k|_p \geq Th$) triggers attack detection by the anomaly detectors, indicating that the grid states (see Table \ref{States considered}) are deviating from their normal values. Using these three components, the defender agent is encouraged to activate protection schemes at the correct times to minimize operating frequency deviations and protect customer equipment from damage. To train the DRL defender agent, we use the \emph{Deep Deterministic Policy Gradient} (DDPG) algorithm~\cite{lillicrap2015continuous}.

DDPG is an off-policy deep reinforcement learning (DRL) algorithm that combines Q-learning with actor-critic methods to learn both value and policy functions. The agent's action space is bounded. After each action is taken at a sampling instance, the defender agent receives a reward based on its effectiveness in maintaining safe grid operations (i.e., keeping the grid frequency within the safe zone ($S$)). This reward is used by the DDPG agent as a performance metric to optimize its actions. During training, the weights and biases of the underlying actor neural networks are updated via backpropagation to maximize reward.

\subsection{Formal Verification of Defender Agent}
\label{NN Verification}

On completion of the training process of the DRL defender agent, it can synthesize activation sequences of protection schemes that can stochastically make grid operation safe against the considered cyber-physical attack vectors. However, such agents may not always be safely deployable in critical smart grid infrastructures due to their probabilistic approach to taking actions. Therefore, in the next phase of our work, we verify the neural network of the trained DRL-defender agent by performing a set-based execution of the network \cite{bak2020improved, tran2020nnv, bak2017hylaa}. The defender agent network is a fully connected feed-forward neural network with ReLU activation functions. We consider the neural network as a function \text{NN}: $\mathbb{R}^{n_i} \rightarrow \mathbb{R}^{n_o}$, where $n_i$ is the number of observations and $n_o$ is the number of actions. The neural network consists of $k$ layers, where each layer $i$ is defined with a weight matrix $W_i$ and a bias vector $b_i$. For an input point $y_0 \in \mathbb{R}^{n_i}$, the neural network computes an output point $y_k \in \mathbb{R}^{n_o}$ as:

\begin{gather}
x^{(1)} = W_1 y_0 + b_1, \quad y_1 = f(x^{(1)}) \notag \\
x^{(2)} = W_2 y_1 + b_2, \quad y_2 = f(x^{(2)}) \notag \\
\vdots \quad \vdots \notag \\
x^{(k)} = W_k y_{k-1} + b_k, \quad y_k = f(x^{(k)}) \label{affine}
\end{gather}

\begin{figure}[!ht]
\includegraphics[width = \columnwidth]{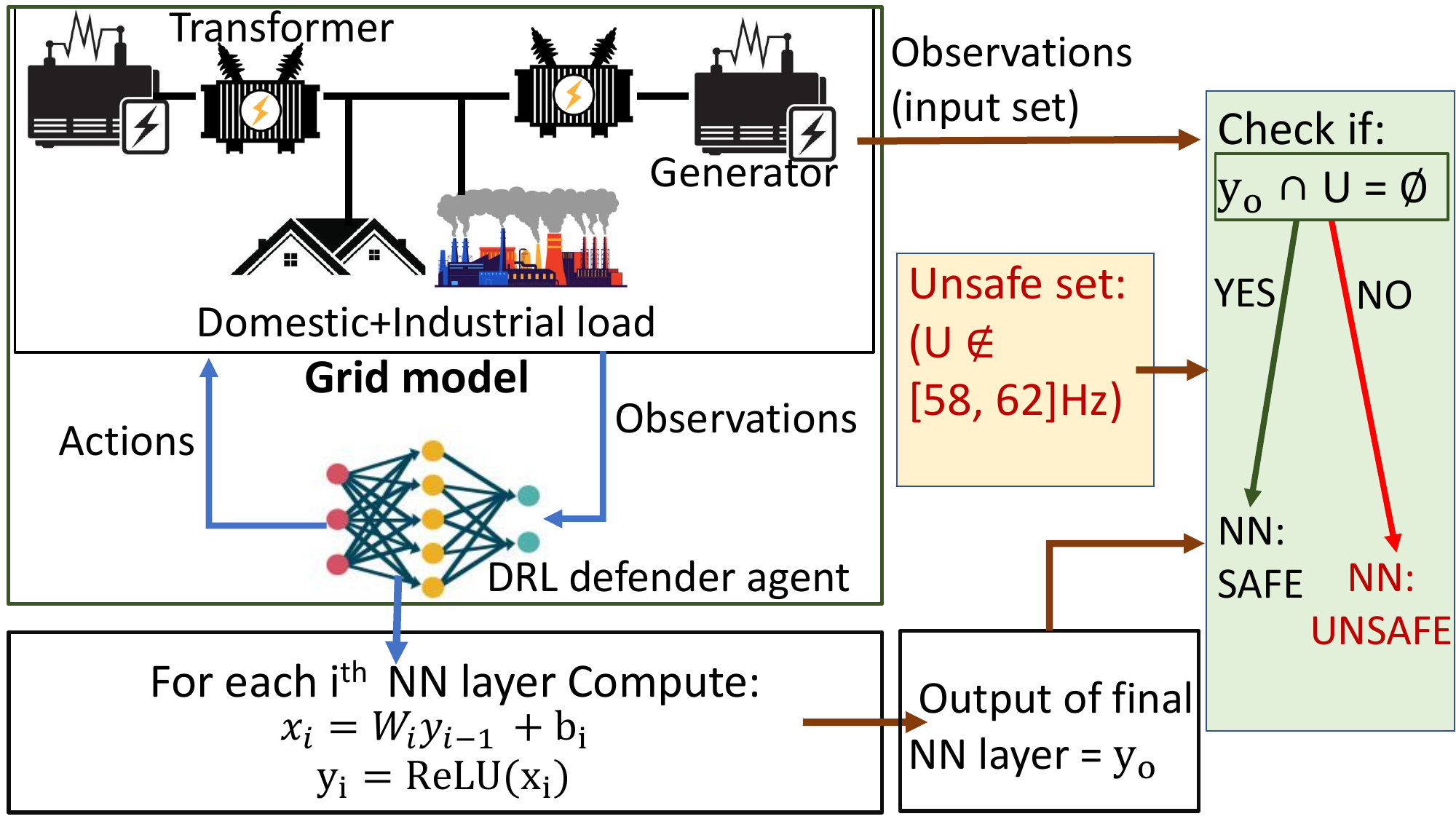}
\caption{Verification framework for the defender agent neural network}
\label{NNV framework}
\end{figure}

In the affine transformation given by Equation \ref{affine}, $y_{i-1}$ and $y_i$ are the input and output of the $i$-th neuron layer, respectively. The function $f$ is the ReLU activation function, where $f(x) = \max(x,0)$. The set-based verification process for the neural networks takes as input the function NN and an input set $I \subseteq \mathbb{R}^{n_i}$ comprising of the agents observations. It then gives an output range ($Range()$) of the network when executed from a point inside the input set, where $Range(NN, I) =\{ y_k | y_k = NN(y_0), y_0 \in I\}$. The verification process checks if $Range(NN, I) \cap U = \emptyset$ for the final output layer of the neural network of the defender agent. Here, $U$ is the unsafe set of grid states ($U \notin [58, 62]$ Hz) indicating potential damage to consumer equipment. Fig. \ref{NNV framework} demonstrates this process. We represent the input set $I$ with the help of a spatial data structure called a star set \cite{tran2020nnv, bak2020improved}. A reachability analysis of the neural network is performed by propagating the input set (observations) through each layer of its neural network, computing the set of possible intermediate values and then the set of possible outputs repeatedly until the output of the final layer is computed and checked for intersection with the unsafe set $U$. The complete process of neural network verification as adopted from \cite{bak2020improved} is presented in Algorithm \ref{DRL verification}.

\begin{algorithm}
\caption{Safety Verification of DRL-based Defender Agent}
\label{DRL verification}
\begin{algorithmic}[1]
\State \textbf{Input:} Input set: $I$, Unsafe set: $U = \{f \mid f \not\in [58, 62]\text{ Hz}\}$, Layers: number of neuron layers, Neuron: number of neurons, Empty pending List: $W$ \label{line1}
\State \textbf{Output:} Status (safe or unsafe) \label{line2}
\State $s \gets \langle \text{layers}, \text{neuron: None}, \theta = \text{star-set}(I) \rangle$ \label{line3}
\If{$s.neuron$ is $None$} \label{line4}
    \If{$s.layer = k$} \Comment{Final layer check} \label{line5}
        \If{$s.output \cap U = \emptyset$} \label{line6}
            \State \Return safe \label{line7}
        \Else \label{line8}
            \State \Return Unsafe  \label{line9}
        \EndIf  \label{line10}
    \Else   \label{line11}
        \State $s.layer \gets s.layer + 1$ \label{line12}
        \State $s \gets s.\text{Affine\_Transformation}(W_{s.layer}, b_{s.layer})$ \label{line13}
        \State $s.neuron \gets 1$ \label{line14}
    \EndIf  \label{line15}
\EndIf   \label{line16}

\State $n \gets s.neuron$  \label{line16}
\State $s.neuron \gets n + 1$  \label{line17}

\If{$s.neuron > \text{length}(b_{s.layer})$} \label{line18}
    \State $s.neuron \gets None$ \Comment{End of current layer} \label{line19}
\EndIf   \label{line20}

\State $activation \gets \text{ReLU\_Activation}(s, n)$ \label{line21}
\If{$activation = \text{positive}$}
    \State \textit{continue} \Comment{No change for positive activation} \label{line22}
\ElsIf{$activation = \text{negative}$} \label{line23}
    \State $s \gets s.\text{reset\_neuron}(n)$ \Comment{Assign 0 value} \label{line24}
\ElsIf{$activation = \text{ambiguous}$} \label{line25}
    \State $t \gets (s.layer, s.neuron, s)$ \Comment{Copy state} \label{line26}
    \State $s \gets s.\text{enforce\_positive}(n)$ \Comment{Positive branch} \label{line27}
    \State $t \gets t.\text{enforce\_negative}(n)$ \Comment{Negative branch} \label{line28}
    \State $t \gets t.\text{reset\_neuron}(n)$ \label{line29}
    \State $W.\text{enqueue}(t)$ \label{line30}
\EndIf \label{line31}
\State $W.\text{enqueue}(s)$ \label{line32}
\State \Return safe \Comment{Continues to be safe} \label{line33}
\end{algorithmic}
\end{algorithm}

The algorithm begins by taking as input the unsafe state \(U\), the input set \(I\) which consists of the observations of the RL agent, the number of neuron layers, and the total number of neurons (line \ref{line1}), as well as an empty list called the pending list. It gives as output the state trajectory of the final layer of the neural network and a status of whether the network is safe or unsafe (line \ref{line2}). The algorithm then forms a computation tuple \(s\), which is a collection of the number of layers of the NN, the affine transformation value of the current neuron being processed (initially set to None), and the star set representation of \(I\) denoted by \(\theta\) (line \ref{line3}). If the current neuron's affine transformation value (\(s.neuron\)) is undefined (None), the algorithm checks if it has reached the final layer (\(s.layer = k\)). If it is the final layer, the algorithm checks whether the current output value ($y_k$) represented as \(s.output\) intersects with the unsafe set (\(U\)). If there is no intersection, it concludes that the NN is safe; otherwise, it returns unsafe (lines \ref{line4}-\ref{line10}). If the neuron is not processed and the current layer is not the final one, the algorithm progresses to the next layer (\(s.layer + 1\)), applies an affine transformation using the weights and biases of this new layer (\(W_{s.layer}, b_{s.layer}\)), and resets the neuron index to 1. This reset indicates the initiation of processing at the new layer, preparing for the application of the activation function (lines \ref{line11}-\ref{line14}). For each neuron in a layer, the algorithm increments the neuron index to move to the next neuron. If the updated neuron index exceeds the number of neurons in the current layer (\(s.neuron > \text{length}(b_{s.layer})\)), it resets the neuron index to None, indicating that it has finished processing the current layer (lines \ref{line17}-\ref{line20}). The algorithm then determines the activation state of the current neuron using the ReLU activation function (\(\text{ReLU\_Activation}(s, n)\)) (line \ref{line21}). Depending on the result of this activation: (i) if the activation is positive, the algorithm continues without making any changes, effectively skipping further checks for this neuron. (ii) If the activation is negative, the algorithm assigns the current neuron's state value to zero (\(s.\text{reset\_neuron}(n)\)) (lines \ref{line22}-\ref{line24}). (iii) If the activation is ambiguous (sometimes positive and sometimes negative), the algorithm branches the current state into two: one assuming a positive outcome and the other a negative outcome (lines \ref{line25}-line\ref{line29}). The algorithm then handles each branch separately as discussed in steps (i) and (ii). Both the branched state (\(t\)) and the original state (\(s\)) are then enqueued into a pending list (\(W\)) (line \ref{line32}) to loop back to line \ref{line6} to check for intersection with the unsafe set \(U\).

The algorithm iterates over this process until the safe/unsafe status of the NN final layer has been determined. We have set a maximum frequency limit of 80Hz for the input star-set $I$, used in Algorithm \ref{DRL verification} to verify the defender agent's NN. This parameter is based on the assumption that the grid operation frequency might rise to 80Hz in the worst-case attack scenarios. Consequently, Algorithm \ref{DRL verification} can verify the defender agent's ability to mitigate such extreme scenarios.

\textbf{Deployment:} The execution of our framework operates as follows. The training of the defender agent is an offline process, where various attack scenarios are utilized to train the agent in learning the optimal protection scheme activation sequences. The trained defender agent's neural network is then formally verified with the help of Algorithm \ref{DRL verification}. After the formal verification process, the defender agent's neural networks are converted to the \emph{dlnnetwork} format \cite{tran2020nnv}, enabling deployment on CUDA-supported GPUs. In the \textit{online} scenario, when the grid is operational, attack scenarios are launched. The defender agent monitors in real-time the operating frequency ($f$) of each generator unit in the grid, the transmission line power flow sensor measurements ($PF$), the residue values of each generator unit ($r_k$), the threshold values of anomaly detection units ($Th$), the circuit breaker control signals ($CB$), and the output power produced by the generators ($P_e$). It successively triggers sequences of protection schemes to safeguard grid operations against the attacks. The defender agent continuously monitors grid parameters and activates appropriate protection strategies to ensure safe grid operation throughout its runtime.

\section{Experimental Setup and Results}
\label{Experiment}

We demonstrate our experimental results for IEEE (14, 37, and 39) bus power grid systems designed using Matlab-Simulink (R2022b) and implemented in Opal RT \emph{Hardware-in-Loop (HIL)} OP4510 setup. These grid architectures are widely accepted models for grid simulations \cite{shekari2022madiot, huang2019not}, where the IEEE 39 bus system represents the New England grid system \cite{shekari2022madiot}. As discussed in sections \ref{attack} and \ref{methodology} we have implemented 5 different attack scenarios, which are: $(i)$ Black IoT \cite{soltan2018blackiot}, $(ii)$ MaD IoT \cite{shekari2022madiot}, $(iii)$ False Data Injection Attack (FDIA) \cite{tan2016optimal, tan2017modeling}, $(iv)$ Scaling Attacks  \cite{maiti2023targeted}, and $(v)$ Circuit Breaker Takeover (CBT) \cite{case2016analysis}, in the Simulink IEEE grid models for training the DRL defender agent. To design and train the defender agent for each power grid model, we use the Reinforcement Learning (RL) toolbox in Matlab. The DRL defender agent was trained using the power grid models along with the attack scenarios (simulated in Simulink) as environments using a high-performance computing setup with a 16-core Intel Xeon CPU with 64 GB of RAM and 16 GB of NVIDIA Quadro 5000 series GPU. We carry out the training process by running 500 episodes, each spanning 50000 iterations with a sampling time of 0.01 s, resulting in a total training time of 9 minutes per episode for each of the bus architectures. In each episode, the defender agent generates optimal activation sequences of the protection schemes to counter each attack scenario in the different grid models. After completing the training of the defender agent, we extracted its Deep Neural Network (DNN), weights ($W$), and biases ($b$) using the \texttt{getActor()} and \texttt{getLearnableParameters()} commands in Matlab. These extracted parameters were then input into a Matlab-compatible Neural Network Verification (NNV) toolbox \cite{tran2020nnv}. The toolbox conducts a reachability analysis of the defender agent neural network (NN) layers, assessing for any overlaps with unsafe grid operation conditions as discussed in algorithm ~\ref{DRL verification}.

\begin{figure*}[ht!]
    \centering
    \begin{minipage}{0.32\textwidth}
        \includegraphics[width=\linewidth]{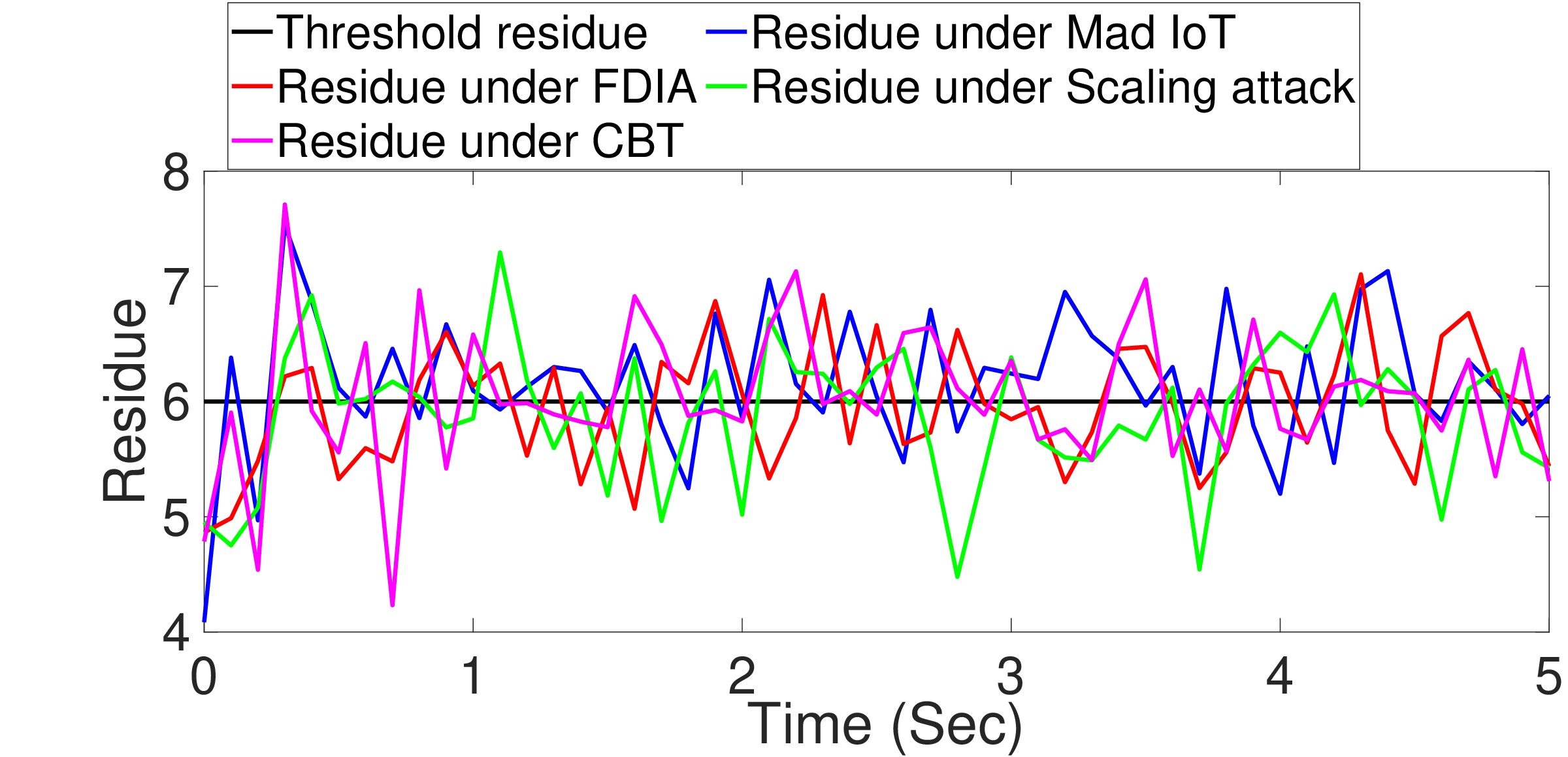}
        \centering
        \caption{ Residue plot of the IEEE 14 bus power grid under different attack models.}
        \label{fig_residue}
    \end{minipage}%
    \hfill
    \begin{minipage}{0.32\textwidth}
        \includegraphics[width=\linewidth]{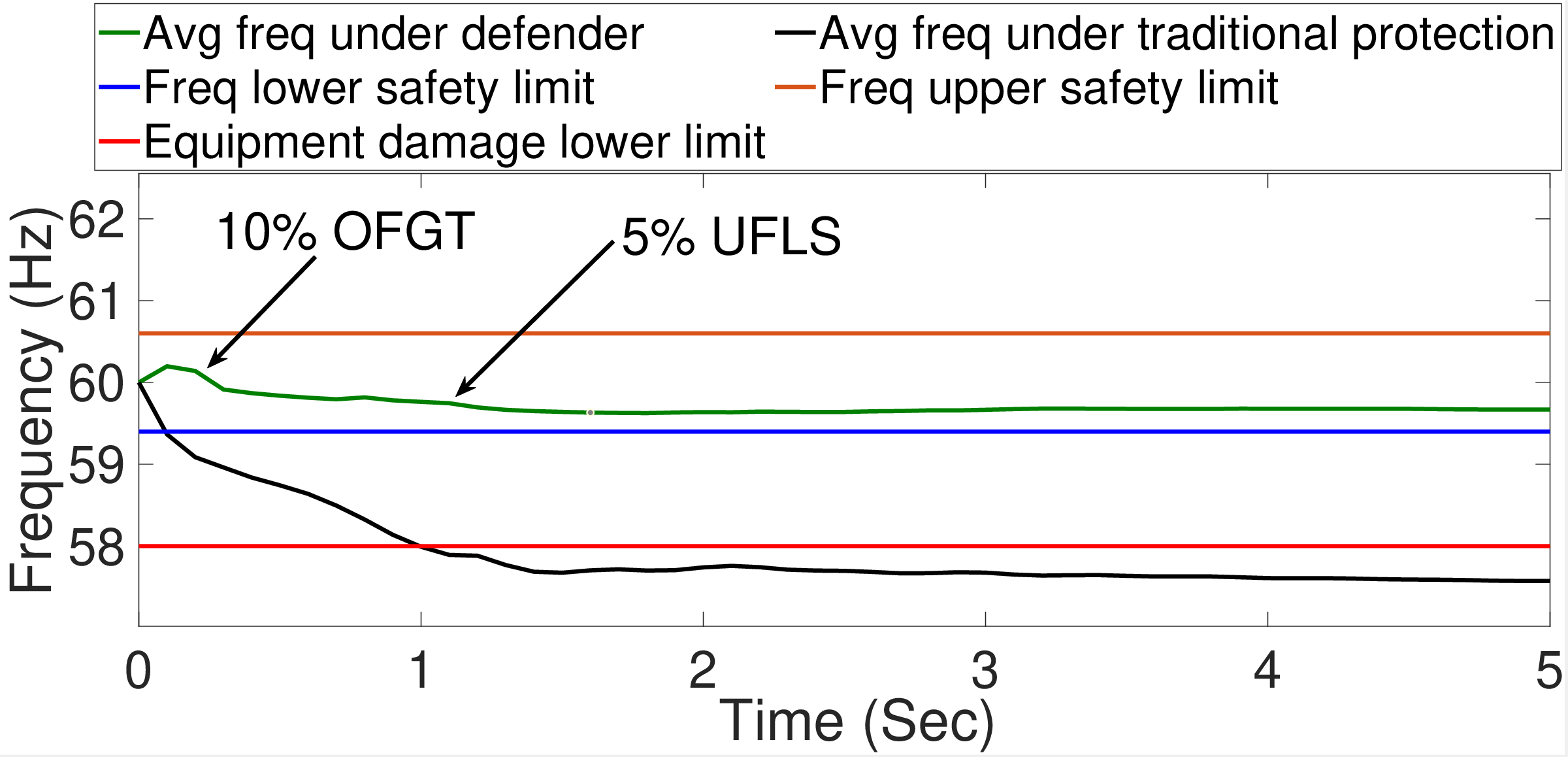}
        \centering
        \caption{ Average frequency of all the generators under MaD IoT attack.}
        \label{fig_freqMAD}
    \end{minipage}%
    \hfill
    \begin{minipage}{0.32\textwidth}
        \includegraphics[width=\linewidth]{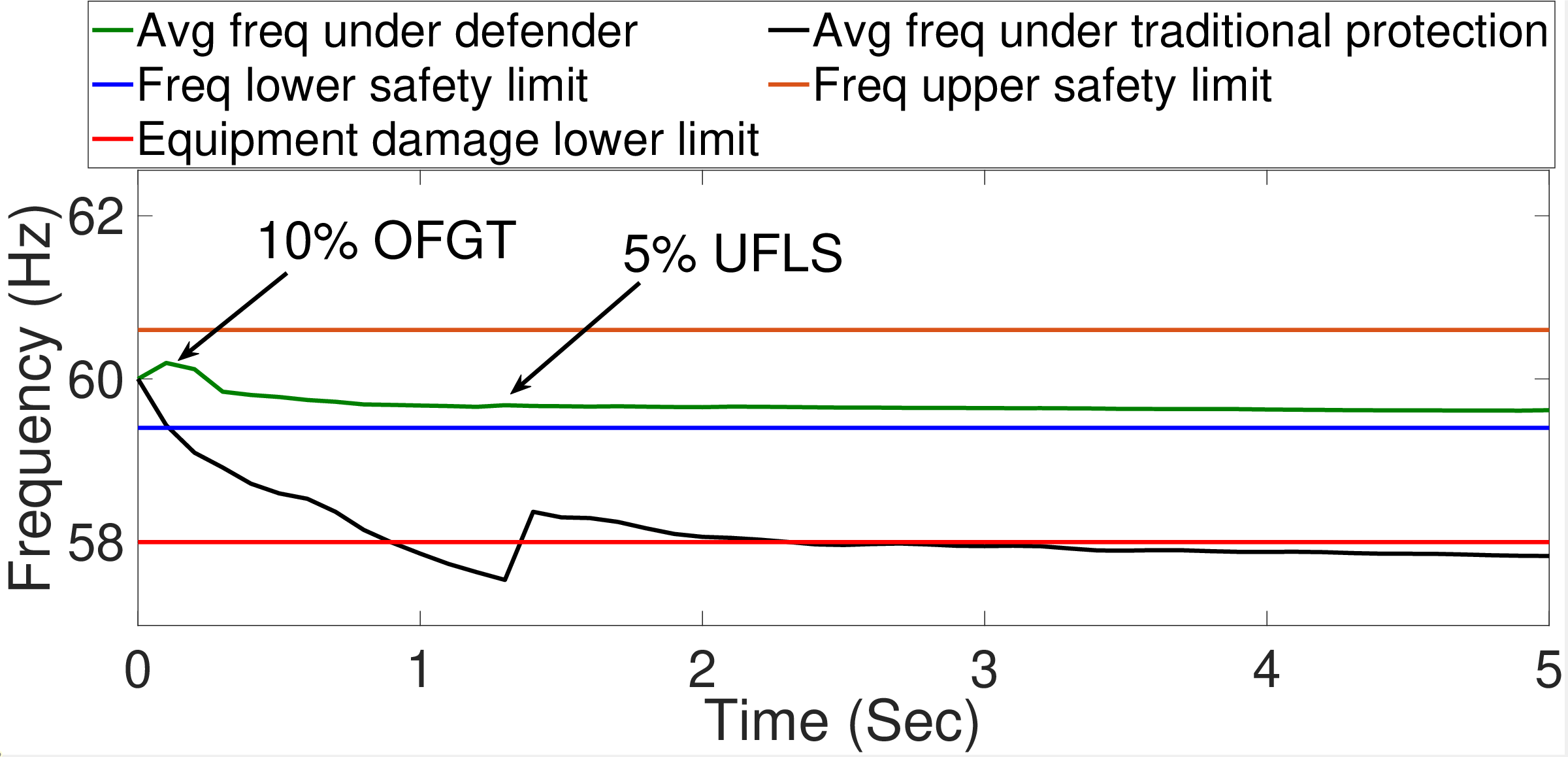}
        \centering
        \caption{Average frequency of all the generators under FDIA attack.}
        \label{fig_freqFDIA}
    \end{minipage}%

    \vspace{1em}

    \begin{minipage}{0.32\textwidth}
        \includegraphics[width=\linewidth]{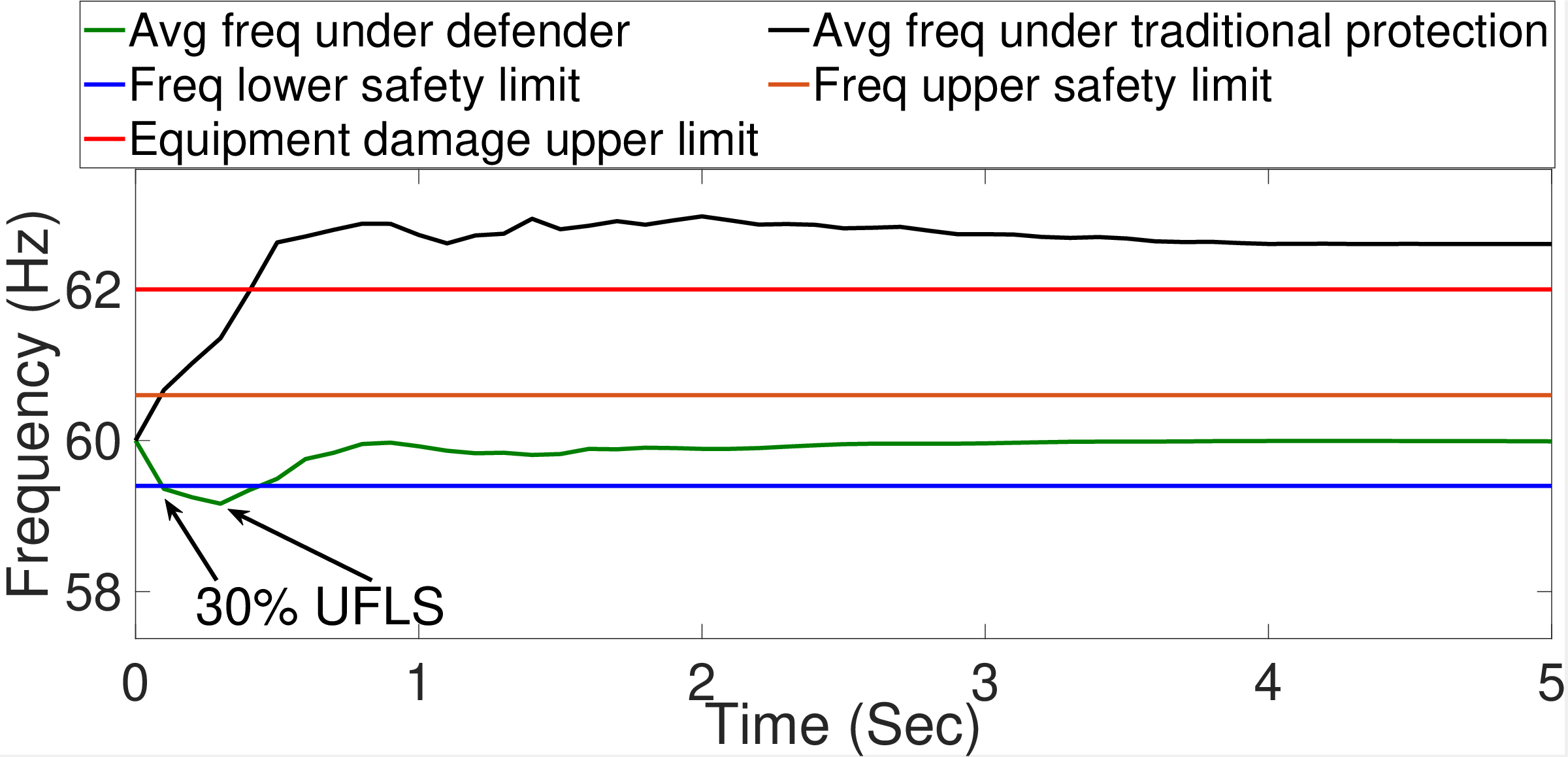}
        \centering
        \caption{Average frequency of all the generators under scaling attack.}
        \label{fig_freqScale}
    \end{minipage}
    \hfill
     \begin{minipage}{0.32\textwidth}
        \includegraphics[width=\linewidth]{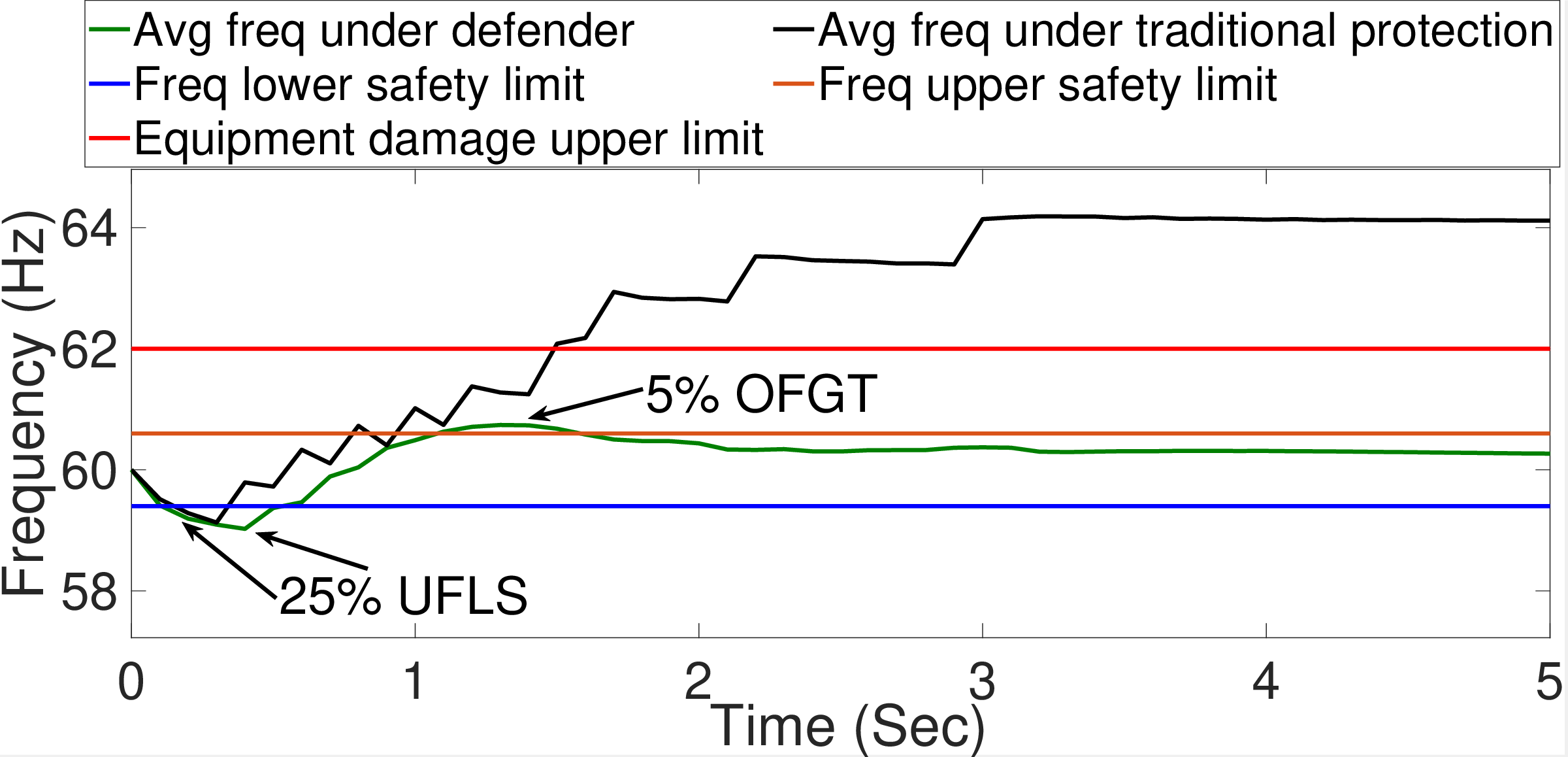}
        \centering
        \caption{Average frequency of all the generators under CBT attack.}
        \label{fig_freqCBT}
    \end{minipage}
    \hfill
     \begin{minipage}{0.32\textwidth}
        \includegraphics[width=\linewidth]{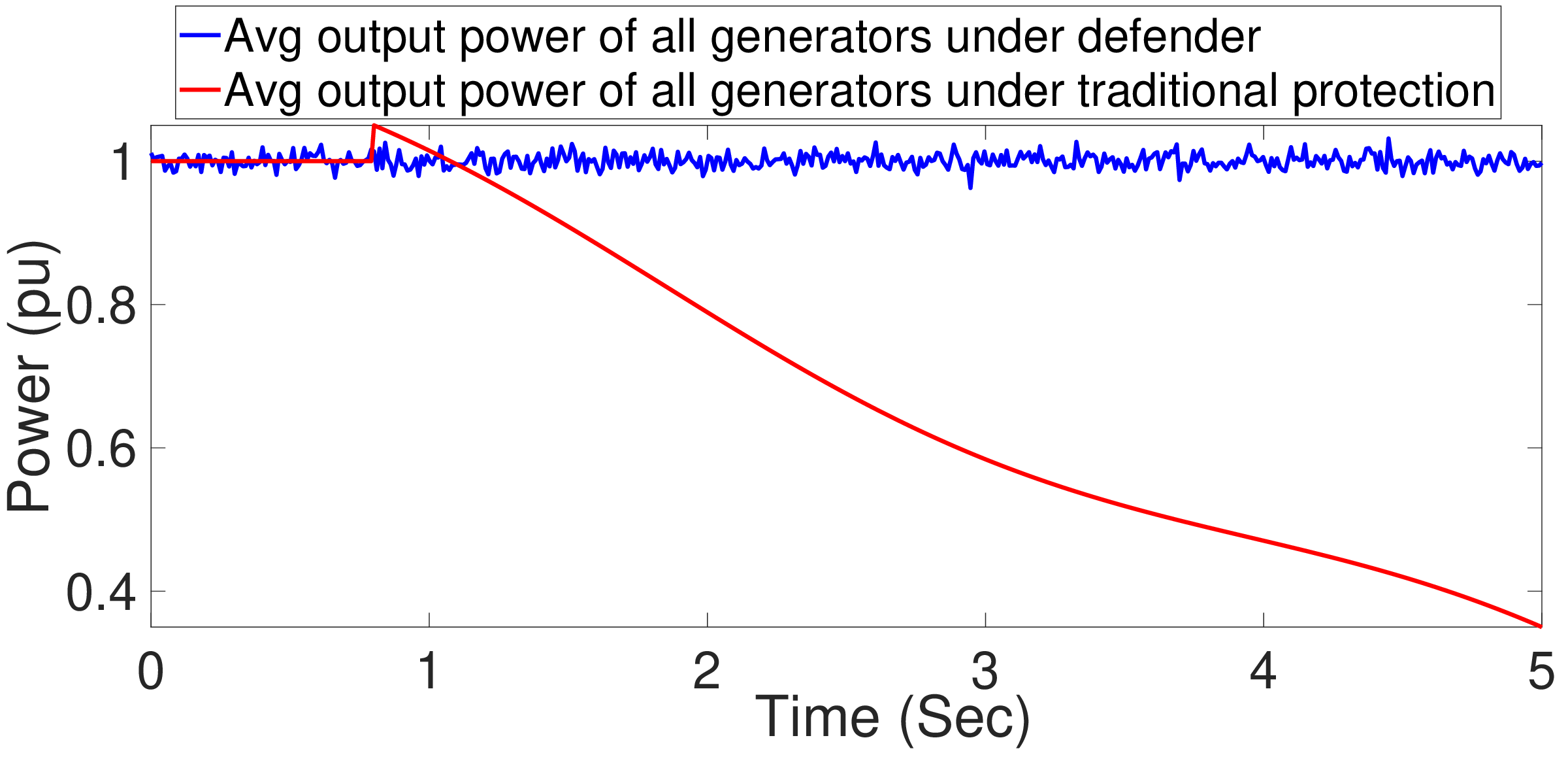}
        \centering
        \caption{Average output power of all the generators under MaD IoT attack.}
        \label{fig_powMAD}
    \end{minipage}

    \vspace{1em}

    \begin{minipage}{0.32\textwidth}
        \includegraphics[width=\linewidth]{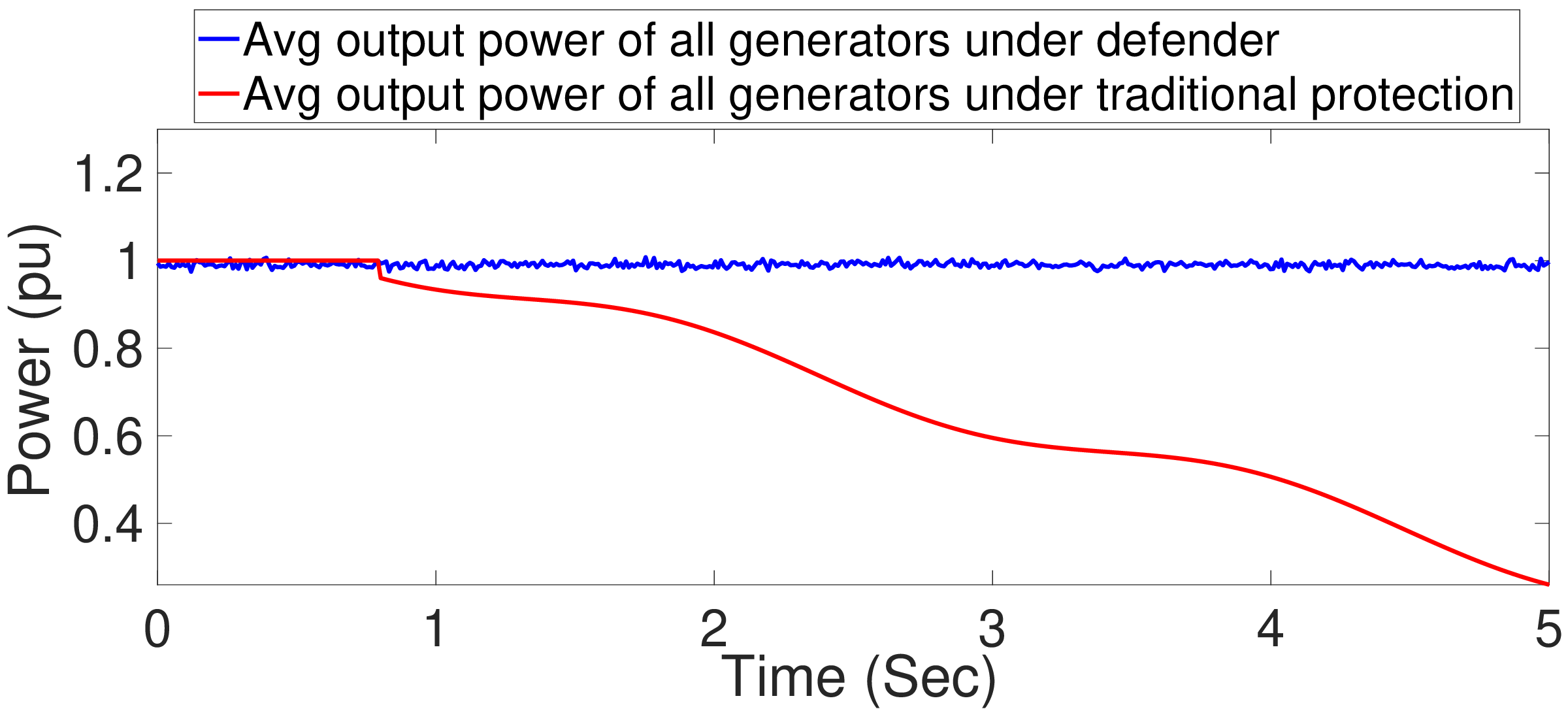}
        \centering
        \caption{Average output power of all the generators under FDIA attack.}
        \label{fig_powFDIA}
    \end{minipage}%
    \hfill
    \begin{minipage}{0.32\textwidth}
        \includegraphics[width=\linewidth]{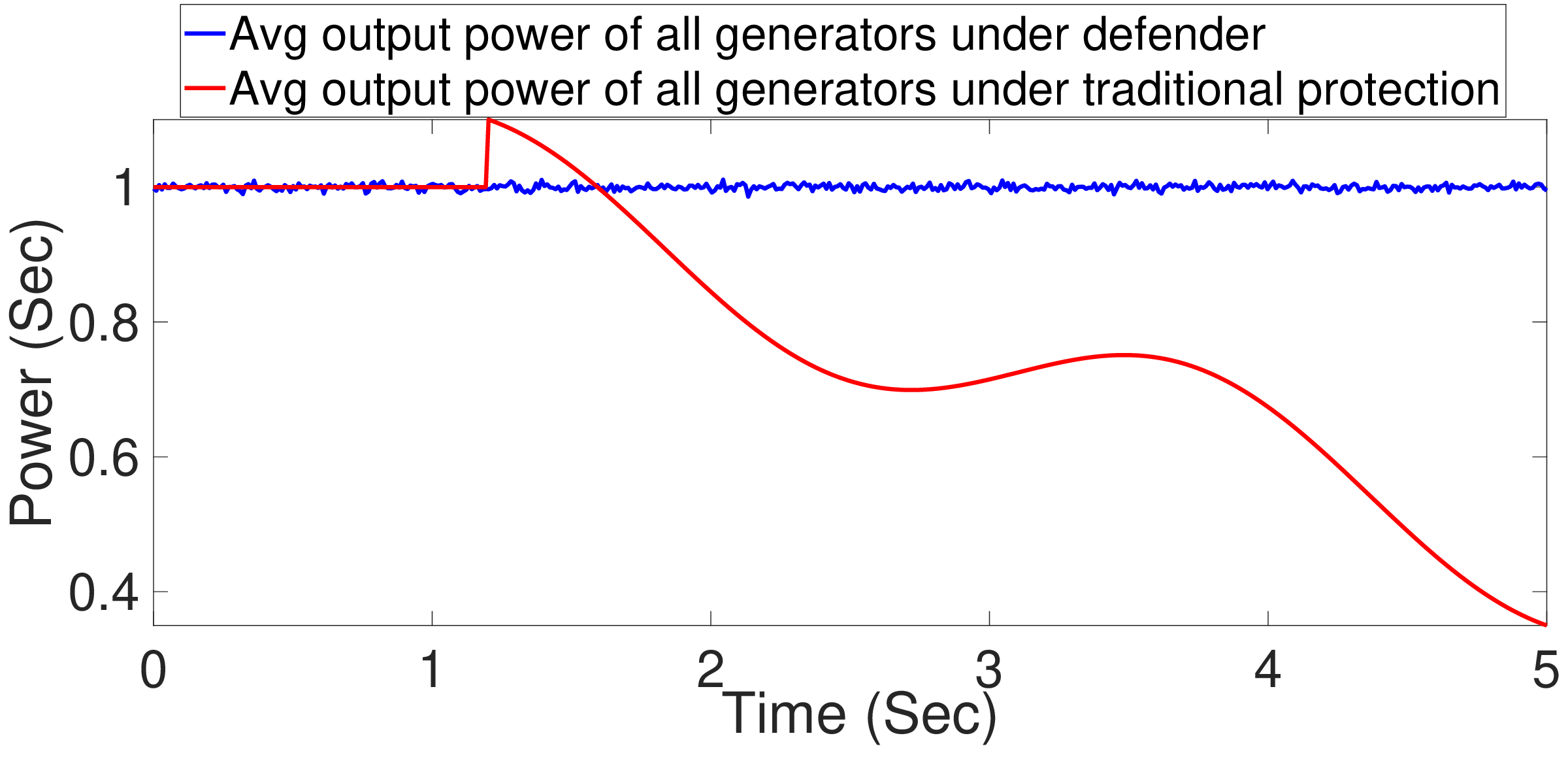}
        \centering
        \caption{Average output power of all the generators under scaling attack.}
        \label{fig_powScale}
    \end{minipage}%
    \hfill
    \begin{minipage}{0.32\textwidth}
        \includegraphics[width=\linewidth]{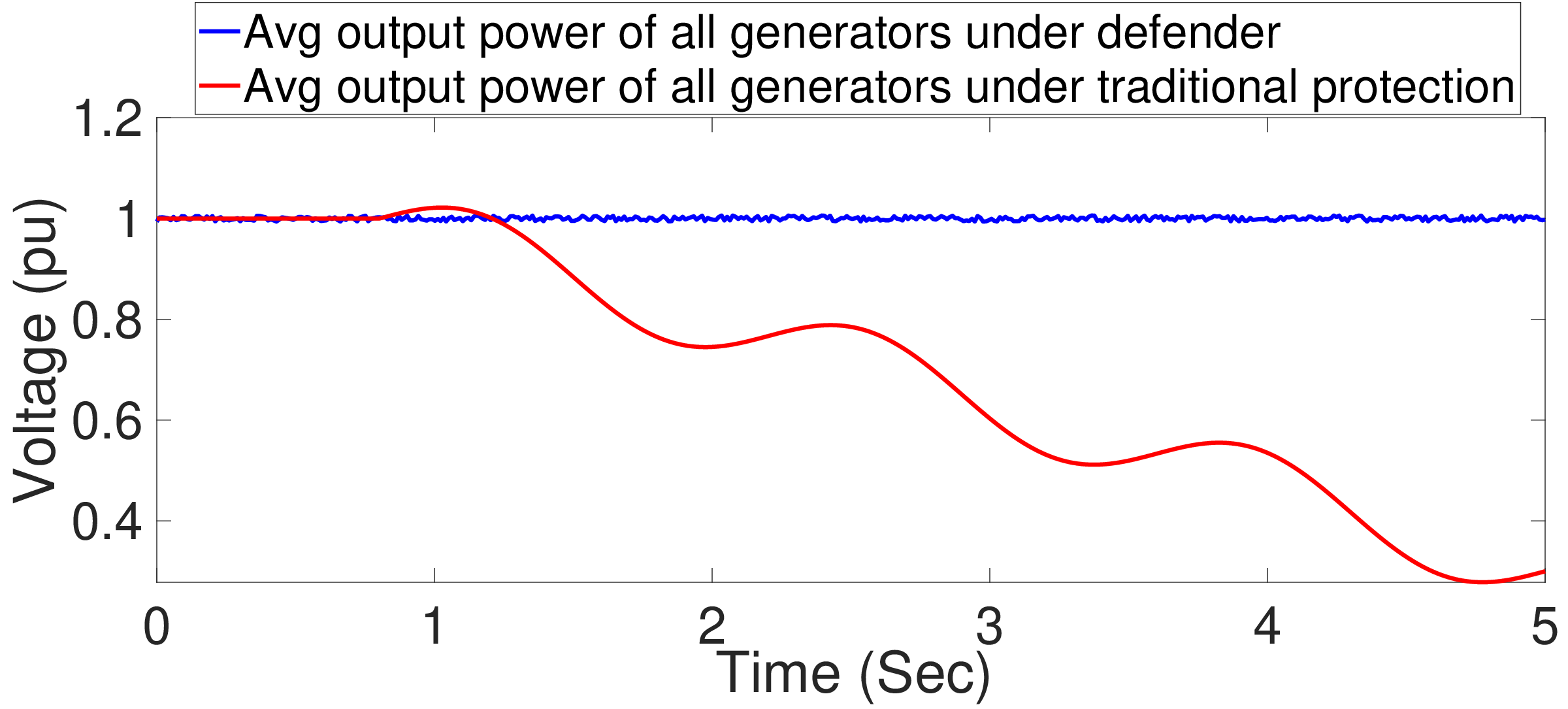}
        \centering
        \caption{Average output power of all the generators under CBT attack.}
        \label{fig_powCBT}
    \end{minipage}
    
\end{figure*}

The verification process using the NNV tool took 20 minutes in our setup. After verification, we converted the DNN's weights (\(W\)) and biases (\(b\)) into GPU arrays (\emph{dlnnetwork}) using Matlab's parallel computing toolbox, facilitating the deployment of the defender agent on CUDA-supported GPU systems. Upon successful completion of the agent's training, verification, and GPU deployment phases, we loaded the IEEE (14, 37, and 39) bus grid models, along with the attack scenario libraries, into the HIL setup. This setup allows for the emulation of the grid models using the OPAL-RT real-time simulation engine, supported by the Artemis solver for advanced real-time simulation. This configuration enables real-time visualization of the impacts of various attack scenarios on grid behavior in Opal-RT. We then loaded the extracted DNN, weights, and biases onto a laptop equipped with a GTX 1650 GPU. An Ethernet connection is established between the laptop and the HIL setup. The grid models and attack scenarios are emulated in the HIL setup, and the observations for the defender agent (discussed in Section~\ref{defender}) are transmitted to the DNN running on the GPU via Ethernet. The DNN processes these inputs and generates appropriate actions, by producing activation sequences for the protection schemes to mitigate the attack scenarios. The actions are generated using Equation~\ref{affine} and sent back to the HIL setup through the Ethernet connection.

The plots in Figures.~\ref{fig_residue}-\ref{fig_powCBT} demonstrate the efficacy of our defender agent in ensuring safe grid operations and preventing damage to customer equipment under various attack scenarios, through real-time simulation of the IEEE 14-bus model. In Fig.\ref{fig_residue} the blue, red, green, and pink plots depict the residue computed by the anomaly detector when MaD IoT, FDIA, scaling, and CBT attacks are executed respectively. As can be seen, our anomaly detection unit efficiently detects all classes of attacks when these residue plots cross the black threshold plot of the detector. The impact of the defender agent in maintaining grid safety across different attack scenarios is further visualized by plotting the following grid parameters.

\par \textit{(i)} \textbf{Average operating frequency of all generators in the grid:} In figures ~\ref{fig_freqMAD}-\ref{fig_freqCBT}, we plot the average operating frequency of all the generators under the defender agent’s action (green plot), the average operating frequency of all the generators under traditional protection schemes (black), the upper (orange) and lower (blue) safety limits of grid frequency, and consumer equipment damage thresholds (upper/lower both in red) \cite{soltan2018blackiot} for an IEEE 14 bus model. These figures illustrate the grid responses for the MaD IoT, FDIA, scaling, and CBT attack scenarios, respectively. Fig.~\ref{fig_freqMAD} explores a worst-case MaD IoT attack scenario where the adversary uses botnets to manipulate 30\% of the grid loads. Traditional protection schemes fail to ensure safe grid operations as the average frequency trajectory of the generators (black plot) falls below the lower safety limit (blue plot) and eventually crosses into the equipment damage lower threshold (red plot). This failure occurs due to a predefined delay in activating existing protection strategies (refer to Table \ref{protection table}) and rigid rules that mandate shedding a fixed amount of load and tripping 100\% of generators. Conversely, our defender agent effectively mitigates the MaD IoT attack and maintains safe operational frequencies, as depicted by the green plot in Fig.~\ref{fig_freqMAD}. Despite the frequency trajectory remaining within the safe zone, the defender agent initiates 10\% OFGT and 5\% load shedding, triggered by the high residue values (Fig.~\ref{fig_residue}) observed during the attack, thus successfully preventing damage to customer equipment.

In Fig.~\ref{fig_freqFDIA}, we analyze an FDIA attack scenario where 40\% of power flow sensor readings are manipulated. Traditional protection measures fail to prevent customer equipment damage as the grid frequency trajectory (black plot) exceeds the threshold for equipment damage (red plot). However, our defender agent successfully maintains the grid frequency (green plot) within the safe operational zone. Similarly, Fig.~\ref{fig_freqScale} demonstrates the impact of scaling attacks on the average operational frequencies of all generators in the IEEE 14 bus grid, considering a scenario where all tie-line power flow sensor measurements are manipulated. Here, traditional protection schemes prove ineffective, allowing the scaling attack to push the frequency trajectory (black plot) beyond the equipment damage upper threshold (red plot). The defender agent effectively mitigates the impact of the scaling attack by implementing 30\% load shedding at 0.3 and 0.6 seconds (green plot), respectively. In Fig.~\ref{fig_freqCBT}, we demonstrate the grid frequency trajectories under a Circuit Breaker Takeover (CBT) attack scenario, assuming a worst-case scenario where the adversary compromises 50\% of circuit breakers. The defender agent counters this attack effectively, maintaining a safe grid frequency (green plot) by initiating 25\% load shedding at 0.4 and 0.7 seconds and 5\% generator tripping at 1.7 seconds. Despite its efforts, traditional protection schemes remain ineffective, with the frequency trajectory (black plot) surpassing the upper safety limit (orange plot) and eventually crossing the upper equipment damage threshold (red plot).

\par \textit{(ii)} \textbf{Average output electrical power of the generator units in the grid:} In figures ~\ref{fig_powMAD}-\ref{fig_powCBT}, we plot the average output power of all the generators under traditional protection schemes (red), and the average output power of all the generators under defender agent operations (blue). These figures illustrate the responses for the MaD IoT, FDIA, scaling, and CBT attack scenarios. Notably, under the operation of our defender agent, the output power (blue plot) stabilizes around a constant per-unit (pu) value of 1, indicating that all generators are effectively supplying power to the loads connected to the grid despite the attacks. In contrast, under traditional protection systems, the power outputs (red plot) exhibit rapid oscillations and decrease to a value of 0.3 pu. These fluctuations result from the alternating increase and decrease of frequencies across all the generators, arising from the inefficient operation of the traditional protection mechanisms. We omit the grid responses for the Black IoT attack in this section as its behavior is similar to that of the MaD IoT attack.

\begin{figure*}[ht!]
    \centering
    \begin{minipage}{0.32\textwidth}
        \includegraphics[width=\linewidth]{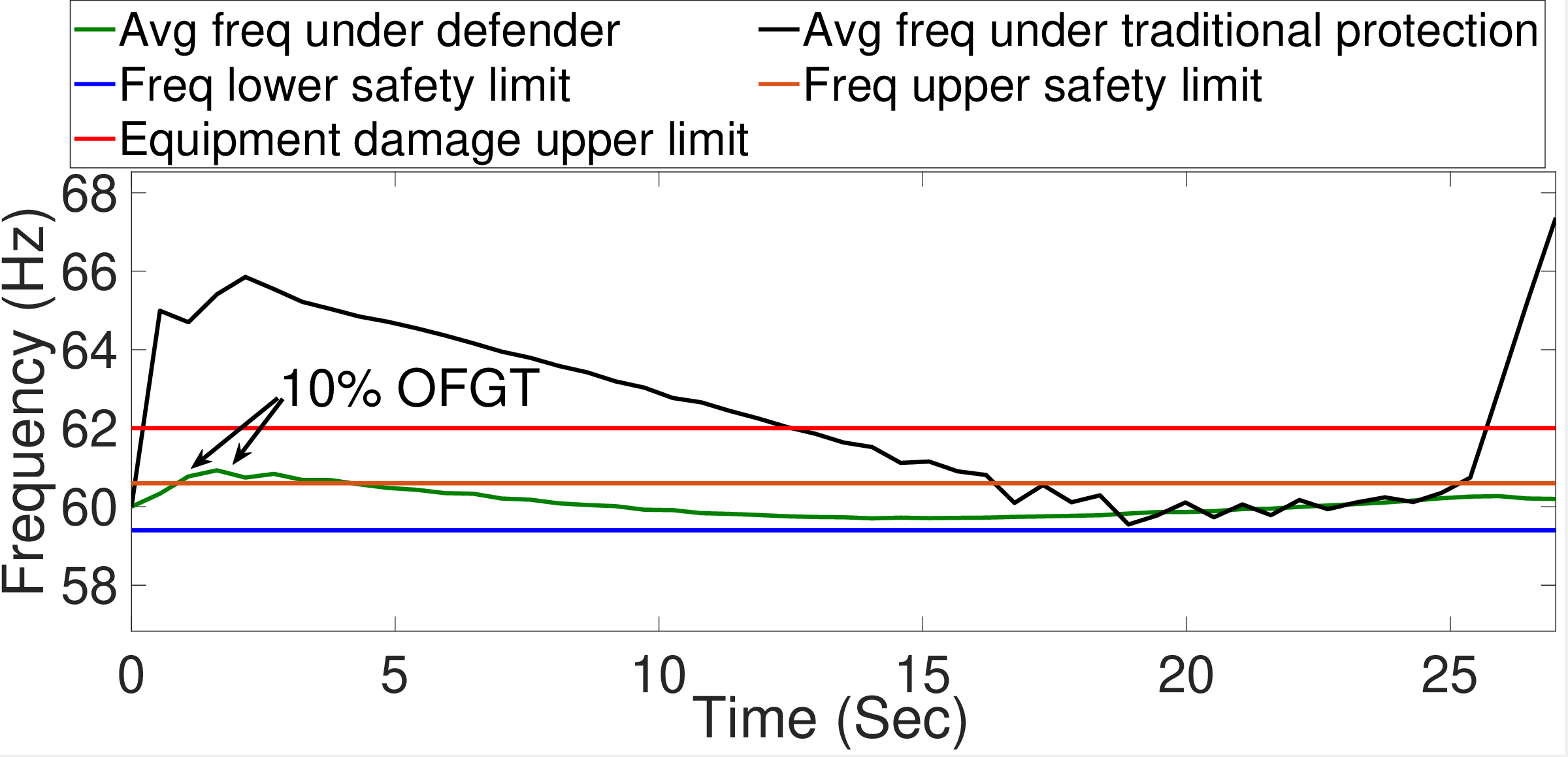}
        \centering
        \caption{ Average frequency of all the generators of IEEE 37 bus under Mad IoT attack}
        \label{fig_freqMad37}
    \end{minipage}%
    \hfill
    \begin{minipage}{0.32\textwidth}
        \includegraphics[width=\linewidth]{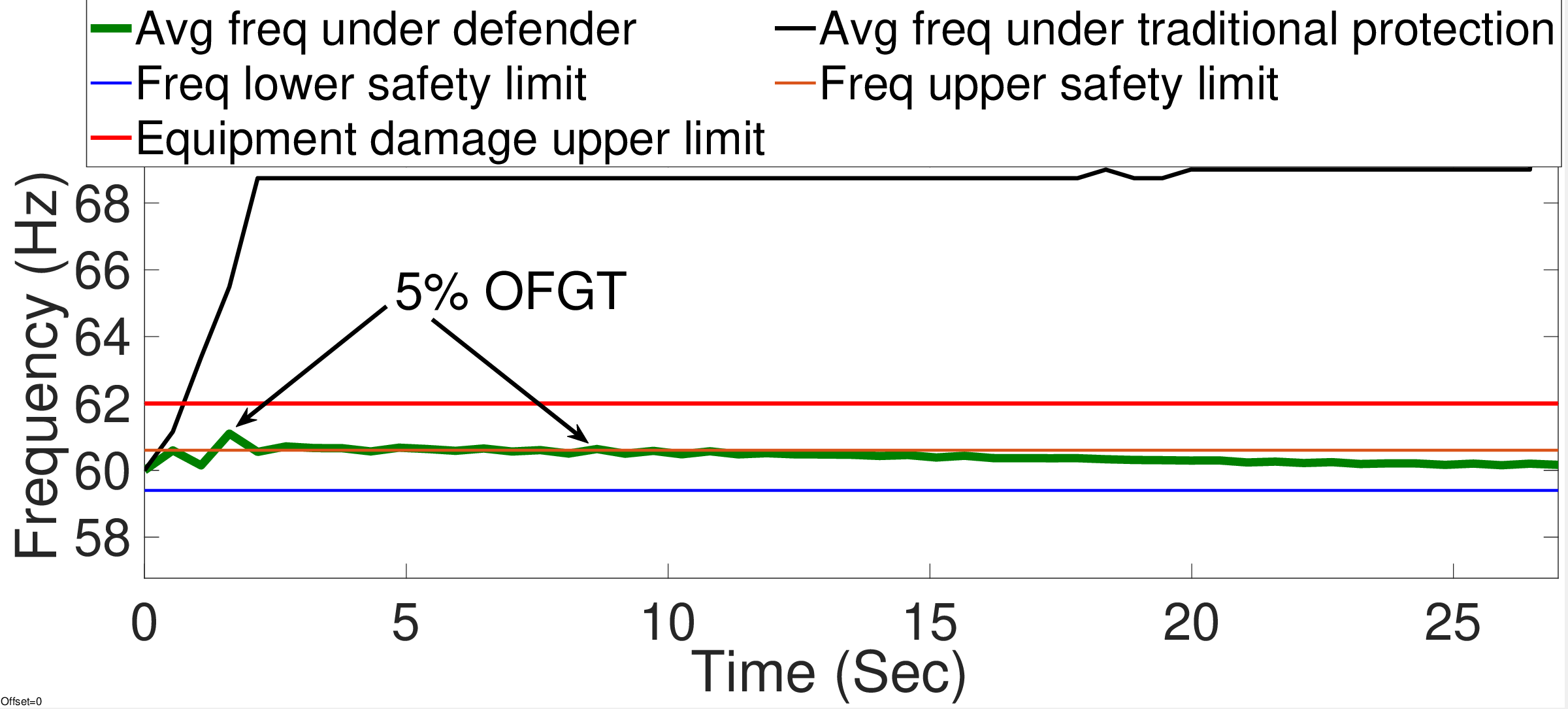}
        \centering
        \caption{ Average frequency of all the generators of IEEE 37 bus under FDIA.}
        \label{fig_freqFDIA37}
    \end{minipage}%
    \hfill
    \begin{minipage}{0.32\textwidth}
        \includegraphics[width=\linewidth]{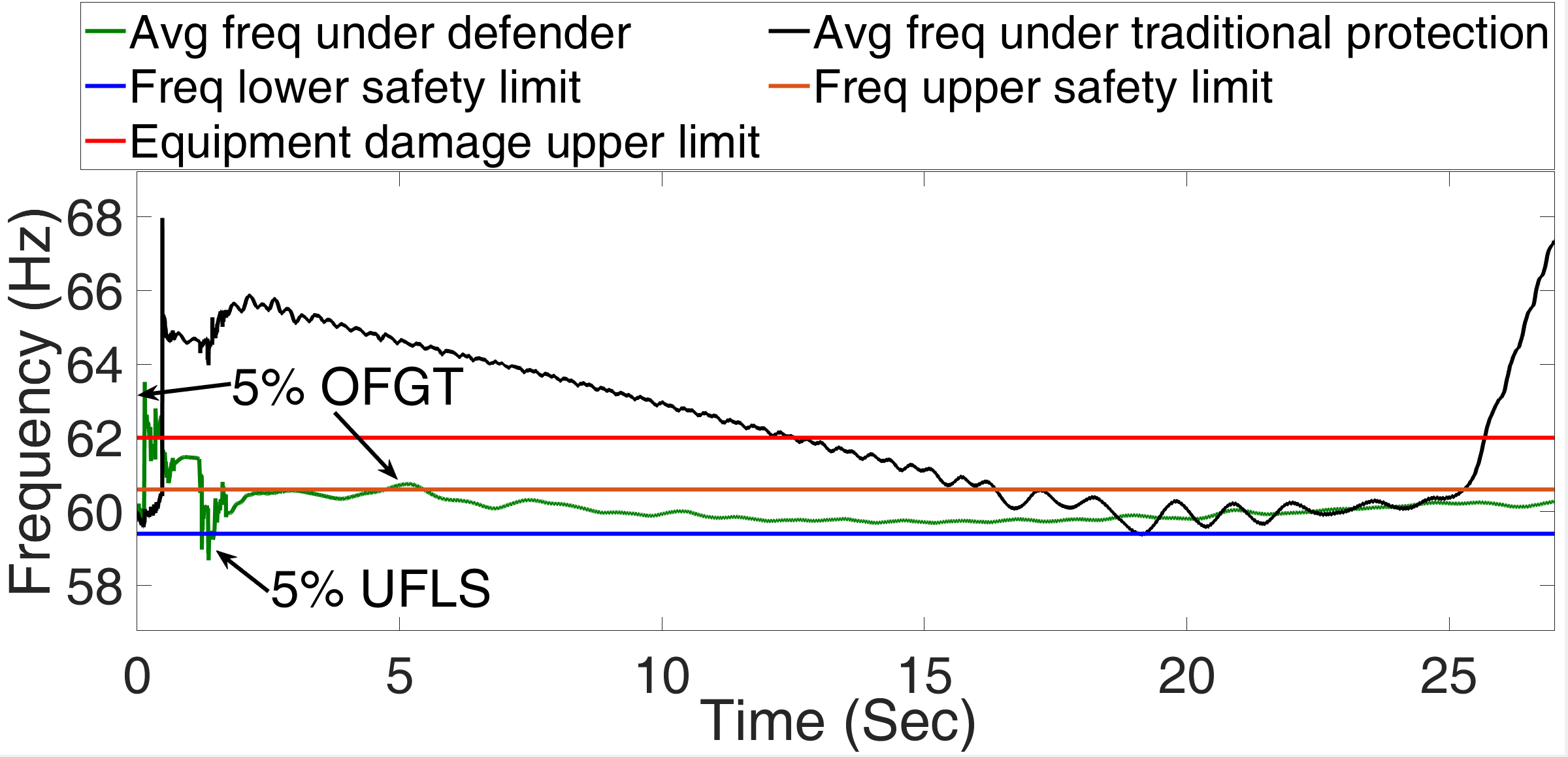}
        \centering
        \caption{Average frequency of all the generators of IEEE 39 bus under Scaling attack.}
        \label{fig_freqScale39}
    \end{minipage}

    \vspace{1em}

    \begin{minipage}{0.32\textwidth}
        \includegraphics[width=\linewidth]{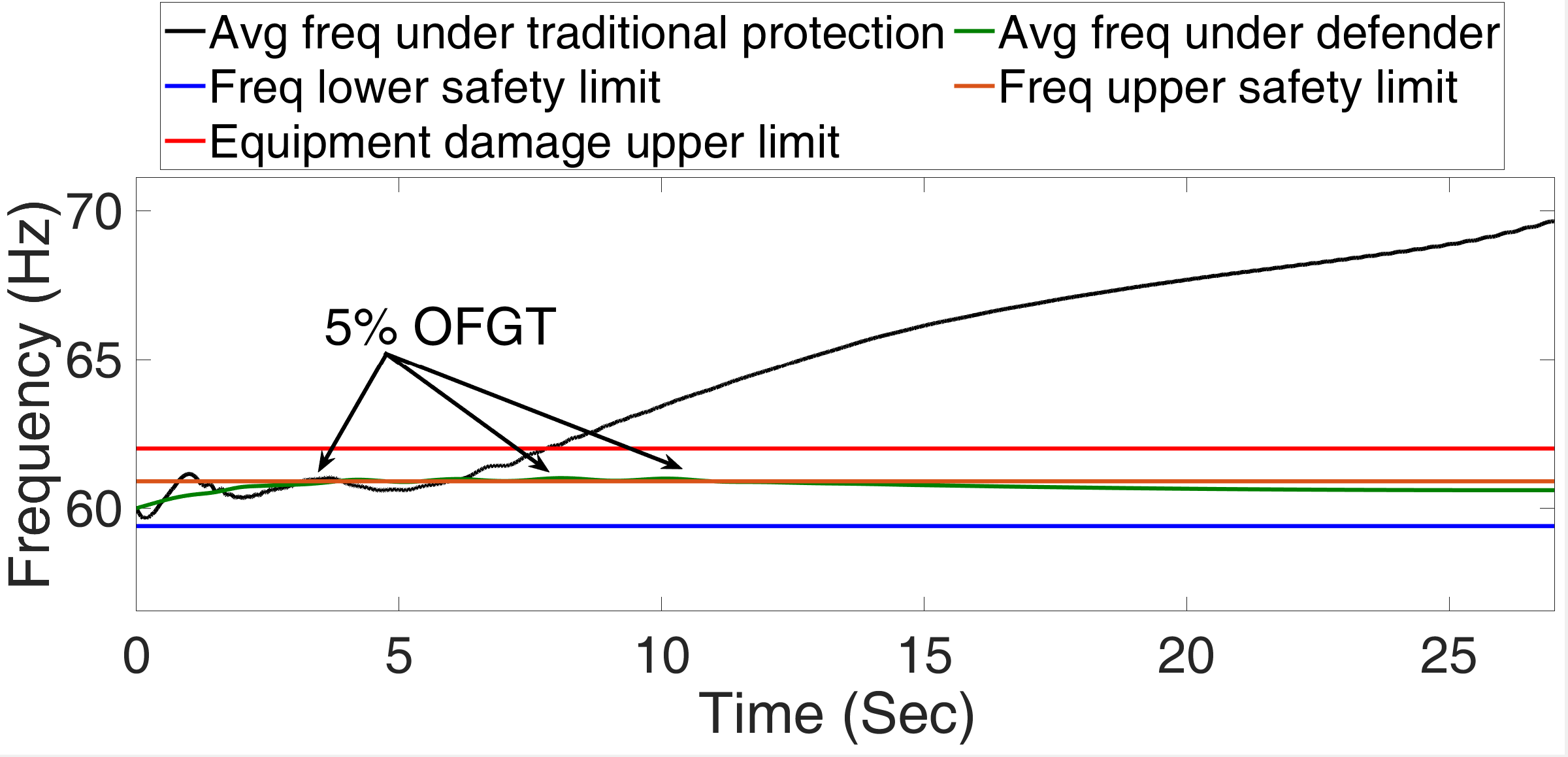}
        \centering
        \caption{Average frequency of all the generators of IEEE 39 bus under CBT attack.}
        \label{fig_freqCBT39}
    \end{minipage}%
    \hfill
    \begin{minipage}{0.32\textwidth}
        \includegraphics[width=\linewidth]{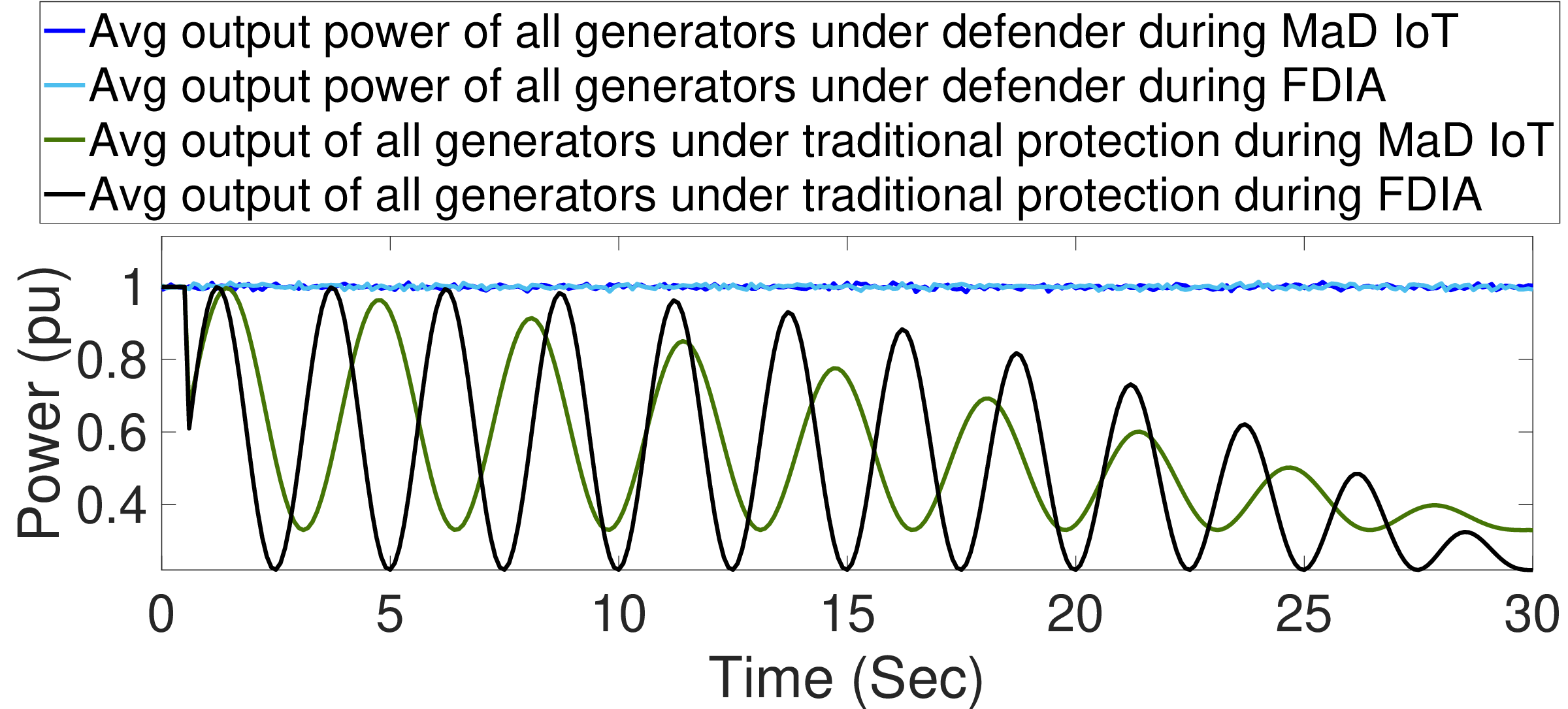}
        \centering
        \caption{Average output power of all the generators of IEEE 37 bus under MaD IoT and FDIA.}
        \label{fig_powMADFDIA37}
    \end{minipage}%
    \hfill
    \begin{minipage}{0.32\textwidth}
        \includegraphics[width=\linewidth]{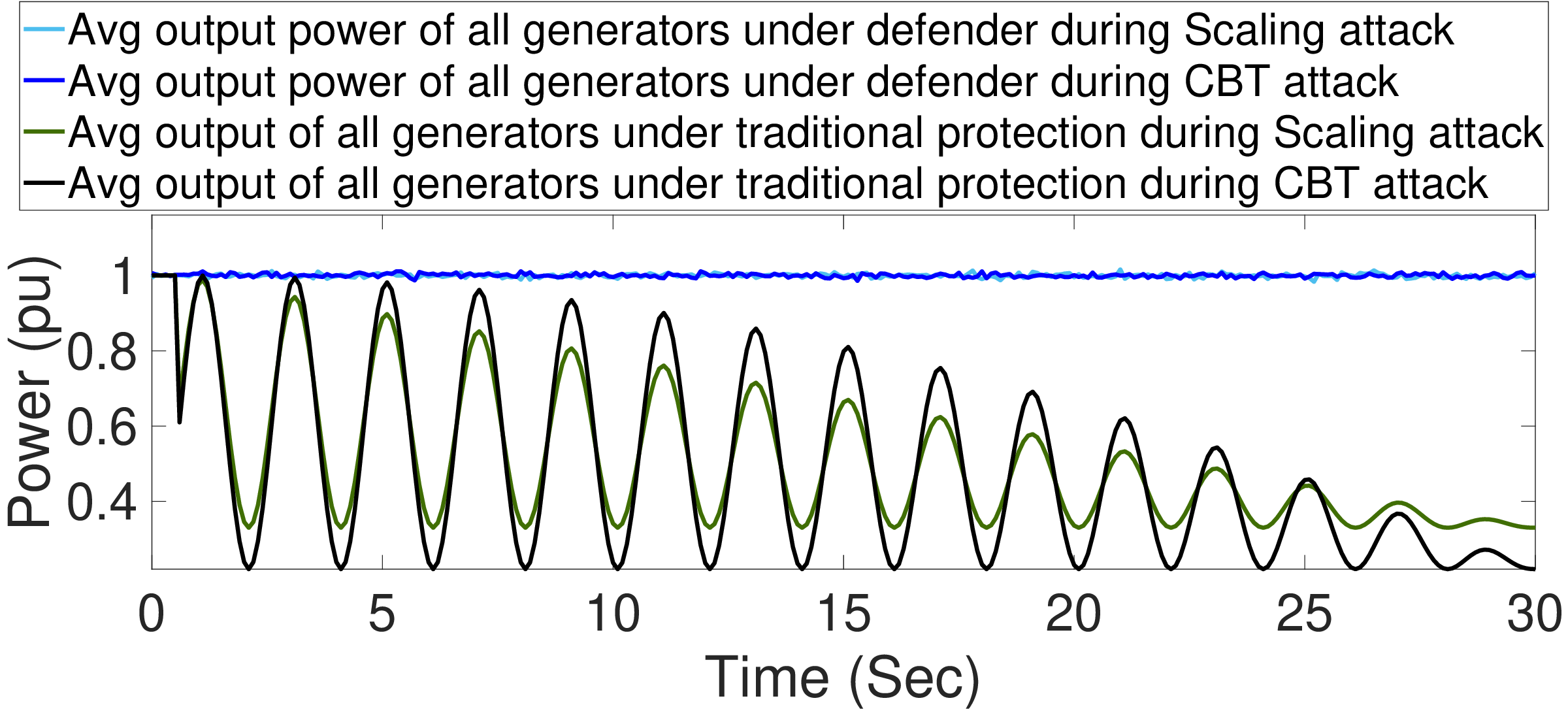}
        \centering
        \caption{Average output power of all the generators of IEEE 39 bus under scaling and CBT attack.}
        \label{fig_freqScaleCBT39}
    \end{minipage}

\end{figure*}

We further plot the average frequency waveform of all the generators in: the IEEE 37 bus model under MaD IoT attack, the IEEE 37 bus model under FDIA, the IEEE 39 bus under scaling attack, and the IEEE 39 bus under CBT attack in figures~\ref{fig_freqMad37}, \ref{fig_freqFDIA37}, \ref{fig_freqScale39}, and \ref{fig_freqCBT39} respectively. The green, black, blue, orange, and red plots in these figures represent the frequency under the operation of the defender agent, traditional protection schemes, lower safety limit, upper safety limit, and equipment damage thresholds (upper/lower both) respectively. Similar to the IEEE 14 bus case, the defender agent successfully mitigated the effects of the different classes of attack on both the IEEE 37 and IEEE 39 bus grid models, ensuring that the grid frequency operated within the safe zone. The traditional protection schemes (Table\ref{protection table}) failed to prevent customer equipment damage in all attack cases.

The average output electrical power of all the generators in the IEEE 37 and IEEE 39 bus grid models is plotted in figures~\ref{fig_powMADFDIA37} and \ref{fig_freqScaleCBT39} respectively, showcasing the MaD IoT, FDIA, scaling, and CBT attack scenarios. In Fig.~\ref{fig_powMADFDIA37}, the plots with colors deep blue and black depict the power output by the generators during the operation of the defender agent and the traditional protection schemes respectively in the MaD IoT attack scenario. The plots with light blue and green colors illustrate the power output of the generators during the operation of the defender agent and traditional protection schemes respectively in the FDIA attack scenario. The efficient operation of the defender agent maintains the power output of the generators at their nominal level even under attack conditions. In Fig.~\ref{fig_freqScaleCBT39}, the plots with colors deep blue and black represent the power output of the generators during the operation of the defender agent and traditional protection schemes, respectively, in the scaling attack scenario for the IEEE 39 bus model. The plots with light blue and green colors represent the power output of the generators during the operation of the defender agent and traditional protection schemes, respectively, in the CBT attack scenario. Similar to Fig.~\ref{fig_powMADFDIA37}, the effective activation of protection strategies by the defender agent ensures that the generators maintain their rated power output even when subjected to scaling and CBT attacks.

The analysis presented through figures~\ref{fig_residue}-\ref{fig_freqScaleCBT39} under various attack scenarios 
 demonstrates the superior performance of our defender agent in ensuring the safety and reliability of grid operations. The real-time simulation results of the IEEE (14, 37, and 39) bus models demonstrate that the defender agent not only mitigates all classes of attacks efficiently but also prevents consumer equipment from damage. This is evident as the defender agent consistently manages to keep the grid's operating frequency and power output within safe thresholds ($S\in [59.5, 60.5]Hz$), a feat that traditional protection schemes fail to achieve.

\subsection{Evaluation of the DRL defender agent}
\label{evaluation}

In this section, we demonstrate the effectiveness of our defender agent in ensuring safe grid operations by comparing it with existing approaches and methodologies.

\noindent \textbf{Strategic Defense Comparison:} We compare the defender with the attack mitigation strategies proposed in \cite{shekari2022madiot} and \cite{maiti2023targeted}. It should be noted that the strategies mentioned in these references are specifically tailored to mitigate MaD IoT and scaling attacks, respectively. Furthermore, the authors acknowledge that these schemes cannot be fully relied upon as there is no mathematical foundation guaranteeing their effective operation under real-world grid operation scenarios. The attack mitigation strategy in \cite{shekari2022madiot} suggests triggering UFLS at the buses where the rate of change of voltage magnitude is the highest. This strategy helps to identify potential buses where the load has been altered via IoT botnets, and successively shedding an appropriate amount of load prevents customer equipment damage. The strategy proposed in \cite{maiti2023targeted} involves triggering 20\% OFGT and 20\% UFGT immediately when the frequency trajectory goes above 60.5Hz and below 59.5Hz, respectively. If the frequency trajectory becomes safe, it reconnects the tripped generators.

\begin{wrapfigure}{r}{0.6\columnwidth}
  \centering
  \includegraphics[width=0.6\columnwidth, clip]{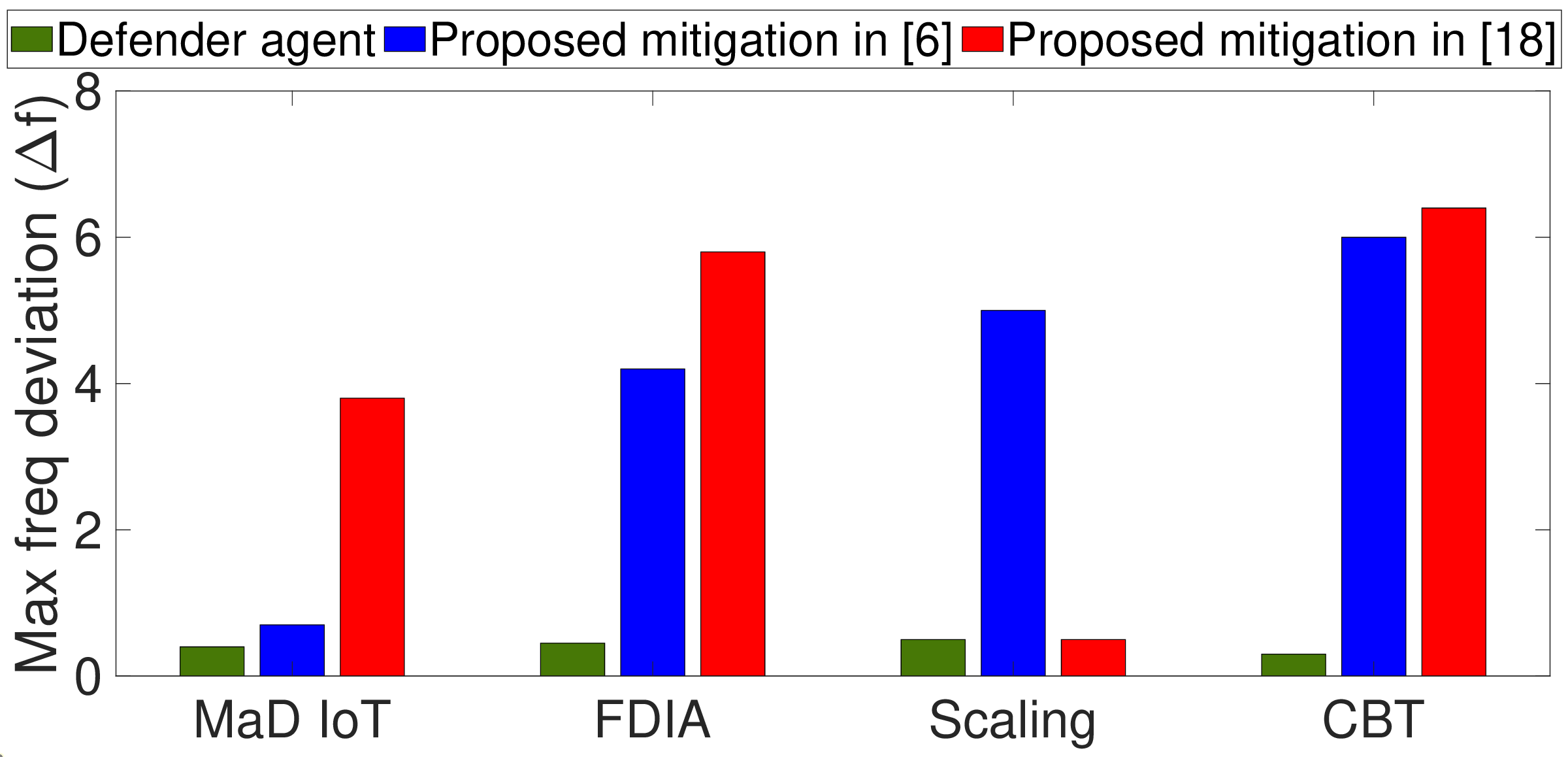}
  \caption{Comparison between different attack mitigation frameworks}
  \label{fig_defComp}
\end{wrapfigure}

This strategy effectively mitigates the scaling attack. The green, blue, and red bar plots in Fig.~\ref{fig_defComp} illustrate the maximum average frequency deviation ($\Delta f$) from 60Hz for all generators in the New England (IEEE 39 bus) grid model. These deviations are measured under the operations of our defender agent, the mitigation strategy proposed in \cite{shekari2022madiot}, and the strategy by \cite{maiti2023targeted}, respectively. Four attack scenarios MaD IoT, FDIA, Scaling, and CBT are depicted in Fig.~\ref{fig_defComp}. For the MaD IoT attack, a total of 10\% of the loads were considered to be altered by the attacker. The FDIA attack was considered to manipulate 25\% of the power flow sensor readings, the scaling attack manipulated three tie-line power flow readings, and 20\% of circuit breakers were considered to be under the influence of an adversary for the CBT attack. It can be seen that the grid operating frequency deviates the least under the operation of our defender agent and remains within the safe frequency deviation limit of 0.5Hz. The other attack mitigation strategies are effective in making grid operation safe for the particular scenarios for which they have been designed, but for other scenarios, they fail to ensure safe grid operation. They are unable to protect customer equipment from damage, as evidenced by frequency deviations exceeding the 2Hz threshold that signals potential equipment damage. Our defender agent outperforms the mitigation strategies outlined in \cite{shekari2022madiot} and \cite{maiti2023targeted}. This advantage stems from incorporating a broad spectrum of attack scenarios during the training phase of our agent. This extensive preparation is complemented by formal verification processes that rigorously evaluate and confirm the efficacy of the defender agent's actions in maintaining safe grid operations.

\noindent \textbf{Performance under natural incidents:} We analyzed the capability of our defender agent in managing natural incidents, such as transmission line outages due to faults and transformer short circuits, and compared its performance 
\begin{wrapfigure}{r}{0.6\columnwidth}
  \centering
  \includegraphics[width=0.6\columnwidth, clip]{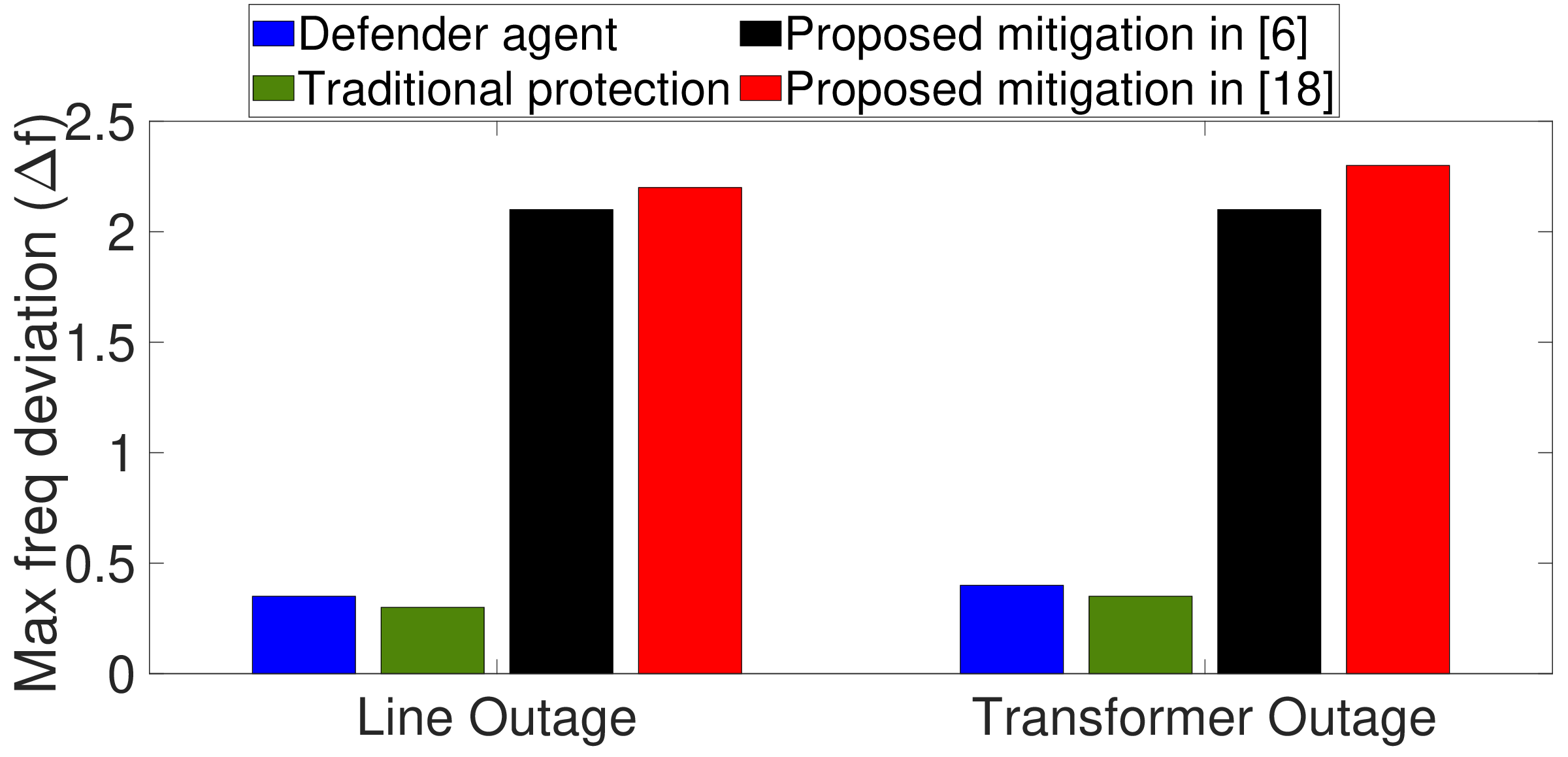}
  \caption{Comparison between defender and traditional protection scheme under natural grid accident scenarios}
  \label{fig_natural}
\end{wrapfigure}
with traditional protection schemes and the protection strategies proposed in \cite{shekari2022madiot, maiti2023targeted}, as illustrated in Fig.~\ref{fig_natural}. The bar plots in blue, green, black, and red depict the maximum average frequency deviation ($\Delta f$) from 60 Hz of all the generators in an IEEE 39 bus grid under the operation of our defender agent, traditional protection schemes, the strategy proposed by \cite{shekari2022madiot}, and the mitigation strategy by \cite{maiti2023targeted}, respectively. For this comparison, we assumed that five transmission lines experienced outages and four transformers were short-circuited. The results indicate that our defender agent performs similarly to traditional protection schemes in handling such natural scenarios, thereby ensuring safe grid operation. However, the strategies of \cite{shekari2022madiot} and \cite{maiti2023targeted} failed to prevent damage to consumer equipment as frequency deviation exceeds 2Hz \cite{soltan2018blackiot}. This efficacy in the performance of our defender agent is attributed to its ability to optimally manage power flow variations caused by these disturbances by effectively solving the power flow Equation.~\ref{mit3}, which forms the basis of its decision-making process.

\begin{wrapfigure}{r}{0.6\columnwidth}
  \centering
  \includegraphics[width=0.6\columnwidth, clip]{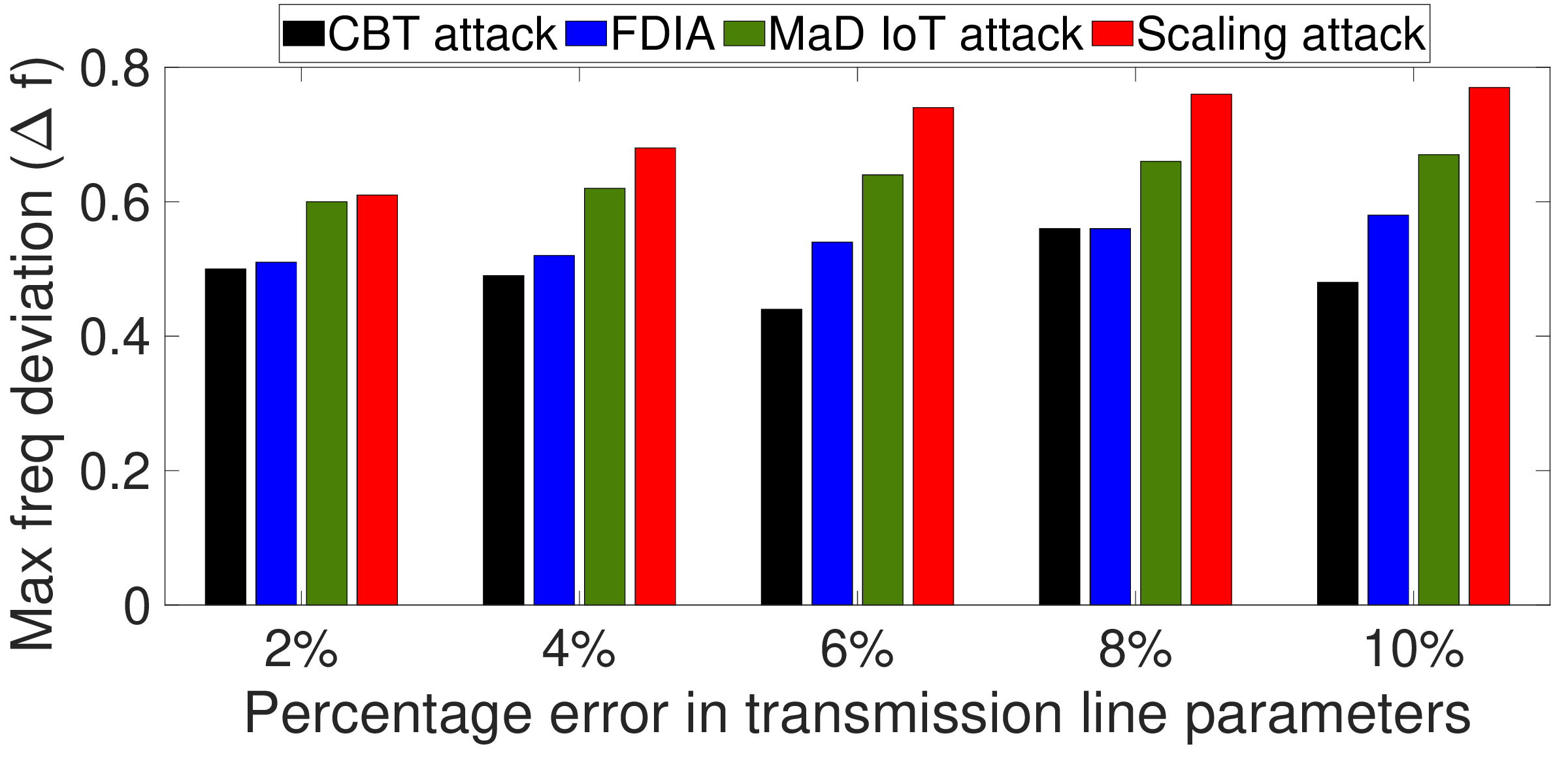}
  \caption{Performance of the defender agent in the presence of errors in grid parameters}
  \label{error}
\end{wrapfigure}

\noindent \textbf{Performance under the presence of error in grid model parameters:} To replicate the real-world uncertainties in modeling the transmission line parameters (reactance, impedance) in our grid setup,  we consider scenarios where certain percentages of errors are present in the reactance and impedance values of the transmission lines within an IEEE 39 bus grid model. We then assess the performance of our defender agent in minimizing grid frequency deviations ($\Delta f$) from 60 Hz across various attack scenarios, as illustrated in Fig.~\ref{error}. The black, blue, green, and red bar plots in Fig.\ref{error} represent the average maximum frequency deviation of all the generators in the grid under the operation of our defender agent during CBT attack, FDIA, MaD IoT attack, and Scaling attack scenarios, respectively.

These attack scenarios have the same adversarial settings as those discussed in Section \ref{Experiment}. As can be seen although the grid frequency deviation exceeds the safety threshold of 0.5Hz due to significant errors in the transmission line parameters, the defender agent successfully prevents customer equipment damage by keeping the frequency deviations below the equipment damage threshold of 2Hz (\cite{soltan2018blackiot}).

\noindent \textbf{Neural Network Verification:} The defender agent is trained with the objective of ensuring safe grid operations by maintaining the grid frequency ($f$) within the designated safe range ($S$) of [59.5, 60.5] Hz. However, due to the stochastic nature of the agent, there were instances during its verification process where algorithm~\ref{DRL verification} could generate counterexamples. In these cases, for certain observations ($I$), the output from the final layer of the trained agent's neural network indicated unsafe conditions, i.e., $f \notin [59.5, 60.5]$ Hz. It is important to note that deviations from the safe range $S$ do not significantly compromise consumer equipment unless the frequency exceeds 62 Hz or drops below 58 Hz, which results in damage to customer equipment, as discussed in \cite{huang2019not, soltan2018blackiot}. As outlined in section~\ref{attack}, the primary goal of the attack models is to damage customer equipment. Therefore, we define an unsafe ($U$) grid operation condition for the defender agent's verification as $f \notin [58, 62]$ Hz. Under this criterion, algorithm~\ref{DRL verification} verified the defender agent, and no counterexamples could be generated as the defender always maintained $f$ within the [58,62] 
Hz range. This criterion ensures that the defender agent's actions effectively prevent damage to customer equipment under all attack scenarios.

With these experimental and comparison results, we demonstrate that our attack mitigation strategy effectively secures grid operations against a vast scenario of sophisticated cyber-physical attacks. Additionally, our strategy proficiently handles natural grid outage events, an aspect that is lacking in the existing research works \cite{shekari2022madiot, maiti2023targeted}. The effectiveness of our proposed method can be attributed to the extensive array of cyber-physical attack models considered during the training of the defender agent, which is followed by its formal verification. Furthermore, our methodology has proven to scale effectively when applied to real-world grid models, such as the IEEE 39-bus New England model.

\section{Related Works}
\label{related}

In recent years, smart grid security has obtained a noticeable amount of attention. Most of the research works in this area focus on vulnerability analysis of the grids and discuss less about discovering robust mechanisms to detect and mitigate the effects of the attack vectors on grid operations. We discuss some related works in this section to understand the gaps in this domain and motivate our contributions.  

\noindent \textbf{Anomaly detection:} Recent works on smart grid security often focus on single state variables such as grid frequency when developing anomaly detection mechanisms \cite{handschin1975bad, li2021dynamic, wu1989detection, li2023convex, amini2017hierarchical, esfahani2010robust}. However, the North American Electric Reliability Corporation (NERC) imposes strict reliability standards on the power grid to ensure stability under a variety of system conditions and potential contingencies \cite{raun2020fiddling}. These standards are designed to prevent cascading failures and have recently been updated to include cybersecurity protections \cite{raun2020fiddling}, though they do not fully mitigate complex threats that exploit grid transient dynamics \cite{huang2018algorithmic}. In contrast, our methodology monitors all state variables of the smart grid, thereby enhancing the detection of a broader range of attacks and improving responsiveness to transient dynamics.

\noindent \textbf{Attack mitigation:} Research studies \cite{abedini2014adaptive, shekari2022madiot, maiti2023targeted} recommend modifications to the operation of existing grid protection schemes to mitigate the impacts of load alteration and scaling attacks. The works in \cite{abedini2014adaptive, shekari2022madiot} proposed triggering Under Frequency Load Shedding (UFLS) at the buses exhibiting the highest rate of voltage change, aiming to address the power supply-demand imbalance caused by MaD IoT attacks. Meanwhile, \cite{maiti2023targeted} suggested activating 20\% OFGT/UFGT immediately when the grid operational frequency deviates from the safe range of $S \in [59.5, 60.5]$ Hz. However, these attack mitigation strategies are tailored for specific scenarios and do not guarantee effective mitigation against a wide variety of cyber-physical attacks targeting grids, as discussed in Section \ref{evaluation}. Additionally, the efficacy of these strategies in ensuring safe grid operation under natural disturbance conditions, such as transformer outages, remains uncertain.

Our formally verified attack mitigation framework effectively counters a range of cyber-physical attacks on grid behavior and has also demonstrated effectiveness against natural grid disturbances. The training of the defender agent across various attack scenarios, followed by the validation of the agent's neural networks, equips it to handle worst-case adversarial attacks even in the presence of errors in grid model parameters.

\section{Limitations}
\label{limitations}

While our DRL-based defender agent demonstrates robust performance in mitigating various cyber-physical attacks, there are inherent limitations in our current methodology that need to be addressed. A significant limitation is our approach to training the defender agent, where we have considered only one attack scenario at a time. This single-scenario approach may not fully capture the complex dynamics of the grid that can occur when multiple attack vectors are active simultaneously. This omission of multivariate attack scenarios could limit the defender agent's ability to generalize its response strategies to complex and coordinated attacks that might target multiple surfaces of the grid simultaneously. Such limitations in the training setup could potentially reduce the efficacy of the defender agent under diverse or evolving attack conditions not previously encountered during its training phase.

\section{Conclusion and Future Work}
\label{future}

In this paper, we have proposed and validated a deep reinforcement learning (DRL)-based framework designed to enhance the security of smart grids against state-of-the-art cyber-physical attacks by updating traditional grid protection schemes. Through extensive real-time simulations, we demonstrated that our DRL defender agent significantly outperforms traditional protection schemes in mitigating sophisticated attacks across various scenarios, including MaD IoT, FDIA, scaling, and CBT attacks. Notably, our formally verified defender agent ensures reliable grid operation and prevents customer equipment from damage. In the future, we intend to train the defender agent by employing combinations of multiple attack scenarios targeting different surfaces of the smart grids. This will broaden the range of applications of our proposed DRL-based defender. We will also explore other reinforcement learning policies, such as Proximal Policy Optimization (PPO) \cite{wu2021coordinated}, TD3 \cite{egbomwan2022twin}, etc. to determine which policy best enables the defender agent to mitigate attacks.  This approach will allow us to prepare a more comprehensive dataset for the training of the defender agent, thereby enhancing its effectiveness and robustness in real-world settings.

\ifCLASSOPTIONcompsoc

\bibliographystyle{ieeetr}
\bibliography{Ref2}

\end{document}